\documentclass[journal, onecolumn, draftclsnofoot, 12pt]{IEEEtran}
\usepackage{cite}
\ifCLASSINFOpdf
  \usepackage[pdftex]{graphicx}
\else
  \usepackage[dvips]{graphicx}
\fi
\usepackage{caption}
\usepackage{xcolor}
\usepackage{subcaption}
\usepackage{amsmath}
\usepackage{ragged2e}
\usepackage{blindtext}
\usepackage{amsfonts}
\usepackage{amssymb}
\usepackage{algorithm,algorithmic}
\usepackage{array}
\usepackage{siunitx}

\makeatletter
\IEEEtriggercmd{\reset@font\normalfont\fontsize{7.9pt}{8.40pt}\selectfont}
\makeatother
\IEEEtriggeratref{1}
\usepackage[subnum]{cases}
\usepackage[nocomma]{optidef}
\DeclareMathSizes{10}{9}{6}{5}
\usepackage{url}
\usepackage{environ}         
\usepackage{etoolbox}        
\newlength{\myl}
\let\origequation=\equation
\let\origendequation=\endequation
\RenewEnviron{equation}{
\settowidth{\myl}{$\BODY$}                       
\origequation
\ifdimcomp{\the\linewidth}{>}{\the\myl}
{\ensuremath{\BODY}}                             
{\resizebox{\linewidth}{!}{\ensuremath{\BODY}}}  
\origendequation
}
\begin{document}

\title{\Large SpaceRIS: LEO Satellite Coverage Maximization in 6G Sub-THz Networks by MAPPO DRL and Whale Optimization}
\author{\normalsize	 Sheikh~Salman~Hassan, Yu~Min~Park,~Yan Kyaw Tun,~\IEEEmembership{\normalsize Member, IEEE},~Walid~Saad,~\IEEEmembership{\normalsize	 Fellow, IEEE}, Zhu Han,~\IEEEmembership{Fellow,~IEEE,}~and~Choong~Seon~Hong,~\IEEEmembership{\normalsize Senior Member, IEEE}
\thanks{Sheikh Salman Hassan, Yu~Min~Park, and Choong Seon Hong  are with the Department of Computer Science and Engineering, Kyung Hee University, Yongin-si, Gyeonggi-do, 17104, Rep. of Korea. E-mails:{\{salman0335, yumin0906, cshong\}@khu.ac.kr}.}
\thanks{Yan Kyaw Tun is with the School of Electrical Engineering and Computer Science, KTH Royal Institute of Technology, Stockholm, Sweden.
E-mail:  {yktun@kth.se.}}
\thanks{Walid Saad is with the Bradley Department of Electrical and Computer Engineering, Virginia Tech, VA, 24061, USA. Email: walids@vt.edu.}
\thanks{Zhu Han is with the Department of Electrical and Computer Engineering at the University of Houston, Houston, TX 77004 USA, and also with the Department of Computer Science and Engineering, Kyung Hee University, Seoul, South Korea, 446-701, USA. E-mail: hanzhu22@gmail.com.}
}
\maketitle
\vspace{-0.7 in}
\begin{abstract}
Satellite systems face a significant challenge in effectively utilizing limited communication resources to meet the demands of ground network traffic, characterized by asymmetrical spatial distribution and time-varying characteristics. Moreover, the coverage range and signal transmission distance of low Earth orbit (LEO) satellites are restricted by notable propagation attenuation, molecular absorption, and space losses in sub-terahertz (THz) frequencies. This paper introduces a novel approach to maximize LEO satellite coverage by leveraging reconfigurable intelligent surfaces (RISs) within 6G sub-THz networks. The optimization objectives encompass enhancing the end-to-end data rate, optimizing satellite-remote user equipment (RUE) associations, data packet routing within satellite constellations, RIS phase shift, and ground base station (GBS) transmit power (i.e., active beamforming). The formulated joint optimization problem poses significant challenges owing to its time-varying environment, non-convex characteristics, and NP-hard complexity. To address these challenges, we propose a block coordinate descent (BCD) algorithm that integrates balanced K-means clustering, multi-agent proximal policy optimization (MAPPO) deep reinforcement learning (DRL), and whale optimization (WOA) techniques. The performance of the proposed approach is demonstrated through comprehensive simulation results, exhibiting its superiority over existing baseline methods in the literature.
\end{abstract}
\IEEEpeerreviewmaketitle
\section{Introduction}
\vspace{-0.1in} 
\subsection{Background \& Motivations}
\vspace{-0.1in}
\IEEEPARstart{B}{y} 2030, vendors and business consulting firms estimate that roughly 100 billion gadgets will be linked in a huge ecosystem. The worldwide count of active Internet of Things (IoT) devices is projected to nearly triple, increasing from 8.74 billion in 2020 to surpass 25.4 billion by 2030, with a remarkable compound annual growth rate exceeding 20\% \cite{Iot_stats}. By using ubiquitous IoT networks, it is feasible to boost efficiency in transportation, health, marine, etc. Massive IoT networks also have the benefit of improving people's quality of life. 
Despite the numerous benefits they offer to industry and human life, the backhaul of ubiquitous and extensive IoT networks poses a challenging problem. Additionally, in certain regions of the world, Internet access is still restricted, i.e., 35.6$\%$ world population lacks access to the Internet \cite{internet_conn.}. 

To address the aforementioned challenges, densely deployed low Earth orbit (LEO) satellites\footnote{Hereinafter, LEO satellite are considered as only satellite unless otherwise stated.} networks are being developed, to offer high-capability backhaul, seamless worldwide connectivity, and considerably more flexible network access services \cite{6gvision}. The progress of sixth-generation (6G) networks is advancing rapidly, with increasing interest in using satellite networks to complement terrestrial networks, i.e., the Starlink initiative by SpaceX \cite{spaceX}. Satellite networks offer advantages, i.e., global coverage, high data rates, low latency, and robustness, making them suitable for applications like broadband internet access, high-definition video streaming, and augmented reality \cite{6G_background}. However, challenges remain, i.e., efficient spectrum management and the need for new protocols and algorithms to support satellite networks' unique characteristics. Despite these challenges, the potential benefits of satellite networks are significant, as they can meet the growing demand for bandwidth and connectivity while providing a reliable and secure platform for various applications \cite{LEO_background}.

Moreover, reconfigurable intelligent surfaces (RISs) have achieved considerable attention as a favorable technology in 6G networks \cite{RIS_directions}. These surfaces consist of numerous sub-wavelength-sized elements that can dynamically manipulate electromagnetic (EM) signals influencing them. Through adaptive modifications of phase, amplitude, and polarization, RISs enable control and optimization of wireless channels \cite{RIS_lit_1}. This capability compensates for large propagation distances, making RISs an attractive choice. Recently proposed RISs can be programmed to modify impinging EM signals, focusing, steering, and enhancing signal power towards the intended user or object. Typically, RISs are constructed as passive reflecting arrays, i.e., intelligent mirror-like structures, comprising multiple almost passive, low-cost, and energy-efficient reflecting elements. These elements enable altering the direction of impinging EM signals \cite{RIS_lit_2}.

Sub-terahertz (sub-THz, i.e., 0.1$\sim$3~THz) communication holds promise for 6G networks because of its potential for high data rates and low latency \cite{sub_THz_lit_1}. Advancements in sub-THz communications worldwide rely on significant developments in environmental sensing and broadcast manipulation to achieve broadband and secure networking at these frequencies \cite{sub_THz_lit_2}. Sub-THz data transmission speeds are considerably higher than millimeter-wave (mmWave, 30$\sim$300~GHz)solutions, exceeding the performance requirements for 6G networks and enabling various applications like holographic transmissions, augmented (AR) and virtual reality (VR), and the tactile Internet. While initial prototypes of low sub-THz point-to-point links are emerging, researchers are increasingly investigating the properties of sub-THz links in various applications \cite{Salman_ICC}.
\vspace{-0.2in} 
\subsection{Contributions}
\vspace{-0.1in}
This study aims to explore the challenges and unexplored research areas in space-RIS. Specifically, we focus on critical aspects that require further investigation, i.e., the relationship between satellite-remote user equipment (RUE), data routing across satellite hops, phase shift manipulation in space-RIS, power allocation for ground base stations (GBS), and practical considerations for technology implementation. For instance, we address the modeling of the actual sub-THz channel in the space environment. Moreover, a significant gap in the existing literature is the absence of a comprehensive and integrated exploration of the potential of multi-RIS-assisted networking in satellite communication. Previous research studies \cite{LEO_Network1, LEO_Network2, LEO_network3, LEO_networking4, LEO_networking5, LEO_networking6, LEO_networking7, LEO_networking8, RIS_lit_1, RIS_lit_2, RIS_lit_3, RIS_lit_4, RIS_lit_5, RIS_L1, RIS_L2, sub_THz_lit_1, sub_THz_lit_2, sub_THz_lit_3, sub_THz_lit_4, sub_THz_lit_5, sub_THz_lit_6, DRL_LIT_2, MADRL_NTN_Lit_1, MADRL_NTN_Lit_2, MADRL_NTN_Lit_3, MADRL_NTN_Lit_4, MADRL_NTN_Lit_5, MADRL_NTN_Lit_6 } have not collectively addressed the optimization of network utility, particularly end-to-end throughput with the space environment that encompasses multiple satellite-based RISs.

The major contribution of this paper is to explore a novel framework for optimizing the utilization of multi-space-RIS with multiple RUEs under the control of the geostationary Earth orbit (GEO) controller. We also introduce a novel and computationally efficient solution mechanism for network management. This mechanism involves employing routing and space-RIS phase shift decision variables as distinct deep reinforcement learning (DRL) agents. By optimizing each decision variable based on its respective constraints, we design separate Markov decision processes (MDPs), thereby enhancing the practicality of our approach compared to prior works that consider either single or multiple agents (i.e., joint MDPs) to address this problem, which is not suitable for learning the optimal values. In summary, our key contributions include:
\begin{itemize}
    \item We present a novel network architecture that utilizes RISs on satellites to serve RUEs. In this context, we quantify the signal-to-noise ratio (SNR) levels of satellites supporting RUE networks by considering RIS beamforming techniques and transmission characteristics, i.e., carrier frequency. Additionally, we propose a practical RISs' beamforming architecture to enhance the coverage range for sub-THz communications.

    \item Building upon this framework, we formulate an optimization problem to maximize the average data rate for all RUEs. This optimization problem involves joint optimization of several factors, namely, satellite-RUE association, data packet routing within the satellite constellation, space-RIS phase shift, and GBS transmit power, which encompasses active beamforming for the sub-THz downlink communication system from the GBS to the RUEs via multi-hop space-RISs. We demonstrate that the formulated problem is mixed-integer non-linear programming (MINLP) and is classified as NP-hard, thus posing significant computational challenges.

    \item To address this complexity, we disintegrate the formulated problem into three sub-problems: (1) satellite-RUE association, (2) data packet routing in satellite constellations and space-RIS phase shift control; and (3) GBS transmit power control. We employ the block coordinate descent (BCD) method to solve each sub-problem individually. Specifically, the satellite-RUE association sub-problem is addressed using a balanced K-means clustering (BKMC) approach, data packet routing, and RIS phase shift sub-problems are solved using a novel multi-agent proximal policy optimization (MAPPO) DRL framework, and the GBS transmit power sub-problem is tackled using the whale optimization algorithm (WOA).

    \item Simulation results illustrate that our proposed algorithms effectively mitigate the propagation attenuation challenges and enhance the satellite coverage range for sub-THz communications. Moreover, our proposed solution successfully solves the NP-hard beamforming problem, especially when space-RIS-empowered sub-THz networks experience multiple hops, outperforming other baseline methods.
\end{itemize}

The paper is organized as follows: Section \ref{lit_review} presents prior art, Section \ref{system_model} introduces the system model and problem formulation, Section \ref{sol_algo} describes the proposed algorithm, and Section \ref{simul} presents the simulation settings and results. Lastly, Section \ref{conc} concludes our work.

\section{Prior Art}
\label{lit_review}
\vspace{-0.1in}
In this section, we discuss satellite-based networking, RIS networks, and sub-THz networks, and introduce multi-agent deep reinforcement learning (MADRL) techniques for 6G networks.
\vspace{-0.25in} 
\subsection{LEO Satellites-based Networks}
\vspace{-0.1in}
Several studies have addressed resource allocation and transmission techniques in satellite networks. Integration of a rental service was proposed in \cite{LEO_Network1} to enhance resource utilization, while \cite{LEO_Network2} introduced a massive MIMO transmission technique with full frequency reuse to overcome instantaneous channel state information (CSI) challenges. In dynamic satellite networks, \cite{LEO_network3} explored the influence of the Doppler effect on downlink NOMA performance and suggested NOMA with OFDM for improved spectrum efficiency. Additionally, a multi-satellite tracking technique for shipborne digital phased arrays was presented in \cite{LEO_networking4}. Grant-free (GF) access schemes were also investigated to dynamically allocate resources among unscheduled resource blocks in satellite networks based on CSI and user priority \cite{LEO_networking5}. Furthermore, an energy consumption minimization problem for LEO satellite networks (LSNs) was formulated in \cite{LEO_networking6}, while our previous works focused on energy-efficient approaches for data computation and resource allocation in space-air-sea (SAS) networks using LEO satellites \cite{LEO_networking7, LEO_networking8}.
\vspace{-0.25in} 
\subsection{RIS-based Networks}
\vspace{-0.1in}
Recently the research community has increasingly focused on RISs due to their potential in enhancing wireless network performance, including coverage, capacity, and energy efficiency. RISs intelligently manipulate the propagation environment, addressing path loss for both mmWave and THz frequencies, mitigating interference, and improving SNR \cite{RIS_lit_2, RIS_lit_3}. The integration of active and passive components in a RIS-aided hybrid wireless network shows promise for cost-effective capacity growth in the future \cite{RIS_lit_4}. An important advantage of RIS lies in its ability to enable fine-grained 3D passive beamforming, allowing individual surface elements to independently reflect incident signals by managing their amplitude and/or phase. This collaborative approach optimizes signal propagation through directional signal enhancement \cite{RIS_lit_5}. For 6G THz communication networks, an analysis of a RIS-assisted MEC-enabled UAV system was conducted to optimize UAV computing capacity, RIS phase shifting, and sub-THz-band assignment \cite{RIS_lit_6}. To fully harness the benefits of RIS, effective signal processing techniques, such as channel estimates, transmission design, and radio localization, are crucial \cite{RIS_L1}. The authors in \cite{RIS_L2} presented an approach for all RIS elements' phase shifts and multi-antenna BS beamforming, by taking into account uplink transmissions from multiple users supported by multiple RISs.
\vspace{-0.25in} 
\subsection{Sub-THz Communication Networks}
\vspace{-0.1in}
The authors presented a framework for enhancing the SNR in a phased array at the GBS through an electronically switchable sub-array. This sub-array adjusts inter-element spacing based on the satellite's elevation angles, accounting for molecular absorption loss, thermal noise from space and Earth, and other noise sources \cite{sub_THz_lit_3}. To tackle the beam search challenge in sub-THz communications, a fast beam search method is introduced, which restricts beam sweep directions by considering previously selected beam combinations for distributed receivers \cite{sub_THz_lit_4}. Data from rooftop satellite base stations operating at 140 GHz demonstrates significant isolation between terrestrial and satellite networks. Moreover, a clear demarcation between ground mobile UEs and co-channel fixed backhaul links indicates robust isolation in these systems. To minimize interference above 100 GHz, it is vital to contain the energy emitted by ground emitters to the horizon, specifically at elevation angles of 15 degrees or lower \cite{sub_THz_lit_5}. Furthermore, the feasibility of using THz-based low-latency communication service to support virtual reality (VR) is explored by the authors \cite{extra_Thz_by_Prof_Walid}.
\vspace{-0.2in} 
\subsection{Multi-Agnet Deep Reinforcement Learning (MADRL) for Networks}
\vspace{-0.1in}
MADRL has gained significant attention as a research paradigm for enabling learning among network entities within interactive network environments \cite{DRL_LIT_2}. Its application has been explored in various scenarios: Designing a flexible satellite payload for mobile terminals in mobile networks to address limited communication range \cite{MADRL_NTN_Lit_2}. Minimizing communication latency in cognitive satellite-UAV networks, particularly in delay-sensitive scenarios with spectrum scarcity \cite{MADRL_NTN_Lit_3} and UAV trajectory designs \cite{MADRL_by_Prof_Walid}. UE and resource assignment in hybrid terrestrial and non-terrestrial networks (NTN), driven by investments in satellite and high-altitude platforms (HAPs) \cite{MADRL_NTN_Lit_4}. Establishing a multi-beam uplink channel assignment scheme in non-terrestrial networks with interference awareness using MADRL \cite{MADRL_NTN_Lit_1}. In wireless networks, a distributively implemented power allocation scheme utilizes model-free DRL for dynamic power allocation. \cite{MADRL_NTN_Lit_5}. Wireless systems' resource management using suggested methods allows agents to produce simultaneous and distributed decisions without knowledge of others' choices \cite{MADRL_NTN_Lit_6}.
\vspace{-0.25in} 
\section{System Model and Problem Formulation}
\label{system_model} 
\vspace{-0.1in}
We begin by introducing the RIS-assisted satellite network architecture for 6G communication, enabling RUE connectivity through satellite in the sub-THz band spectrum. Next, we present the satellite constellation's structure, encompassing the orbit model and orbital duration. Lastly, we discuss the communication channel models and link analysis for satellite-based communication.
\vspace{-0.5in} 
\subsection{Network Model}
\begin{figure*}[t]
\centering 
\includegraphics[width=0.65\textwidth, height=2.8in]{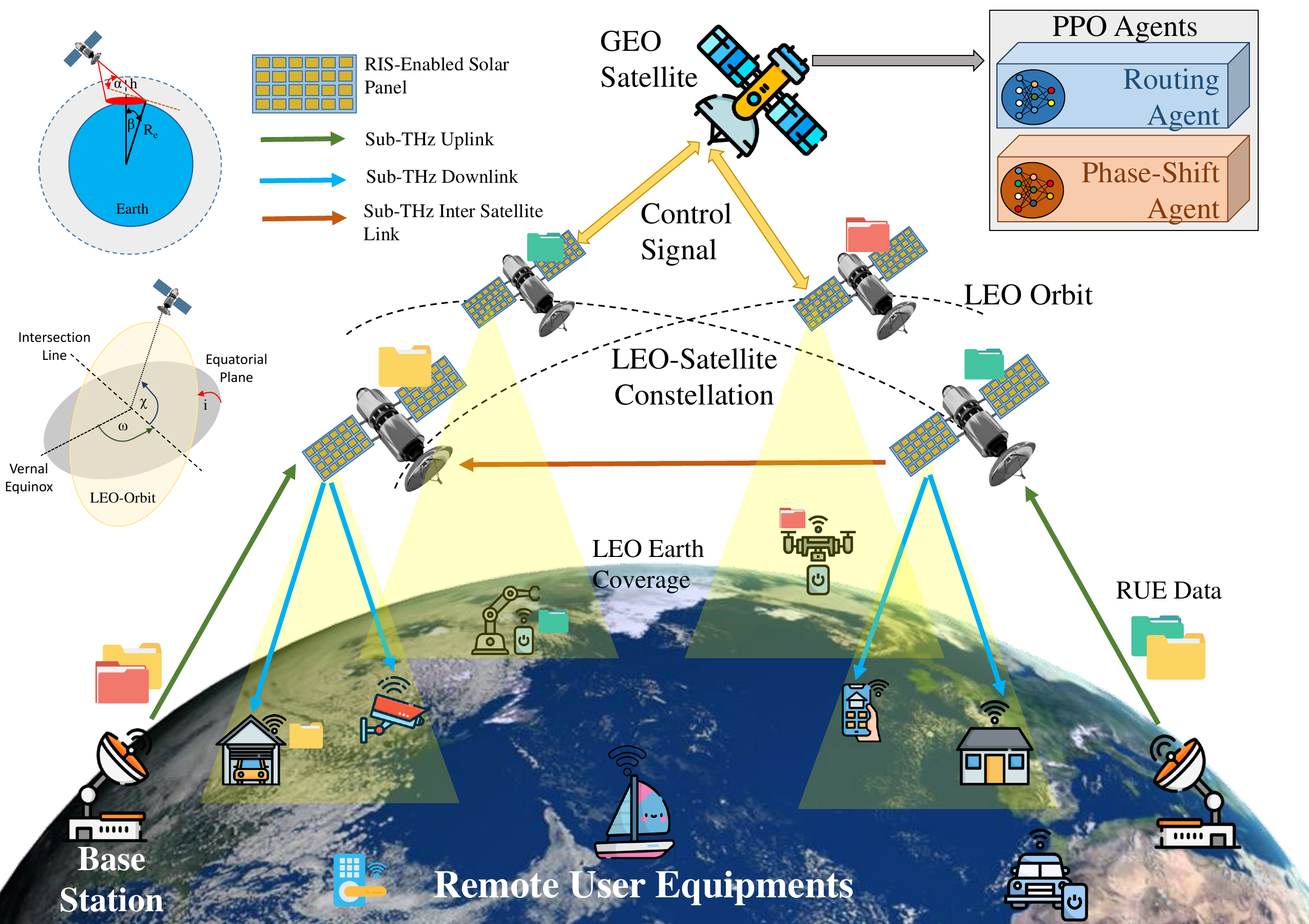}
\caption{Illustration of proposed RIS-assisted LEO satellites' coverage maximization framework for 6G networks.}
\label{sysmodel}
\vspace{-0.4in}
\end{figure*}

\vspace{-0.1in}
We assume a multi-hop RIS-assisted satellite-based multiple-input multiple-output (MIMO) wireless transmission network to address the propagation attenuations at sub-THz band frequencies for RUEs as shown in Fig. \ref{sysmodel}. The network comprises a single GEO satellite (acts as the controller for LEO satellites), a set $\mathcal{S}$ of $S$ LEO satellites, a set $\mathcal{R}$ of $R$ RISs deployed on satellites, and a set $\mathcal{U}$ of $U$ RUEs dispersed following a homogeneous Poisson point process (HPPP). Each RUE $u$ has a respective antenna and each RIS $r \in \mathcal{R}$ composed of a matrix $\boldsymbol{N}_r$ with $N_r$ reflecting elements equipped with phase shifters to combat propagation loss. We also consider a set $\mathcal{B}$ of $B$ GBS to assist the satellites. The network operates within a specific period $T$, separated into a set $\mathcal{T}$ of $T-1$ time slots, during which the network configuration remains fixed. The positions of each GBS $b$ at each time slot $t$ are given as $
\boldsymbol{d}_{b}(t)=\big[\left(x_{b}(t), y_{b}(t) \right)\big]^{T}, ~\forall b \in \mathcal{B}, ~\forall t \in \mathcal{T}.$
Similarly, the position of each RUE $u$ will be: $
\boldsymbol{d}_{u}(t) = \big[x_{u}(t), y_{u}(t)\big]^T, ~\forall u \in \mathcal{U}, ~\forall t \in \mathcal{T}. $
To increase received power levels, we propose using RISs at the receiver end. These meta-surfaces can be modified instantaneously to change reflection phases \cite{RIS_1}. In our model, RISs are placed on satellites instead of reflecting arrays. By adjusting the phase shift for individual meta-atom, the incident signal can be beamformed to desired RUEs. The satellite constellation is structured to guarantee that at least one satellite passes above the area of interest (AOI) at each time slot, providing seamless coverage to RUEs. We consider satellites operate in $M$ orbital planes of altitude $h_s$ and are organized into sets $\mathcal{M}$ and $\mathcal{S}_m$, with each orbital plane $m$ containing $S_m$ satellites. Each satellite is associated with all RUEs in its coverage region, and the RUEs evenly divide the available bandwidth $B$ over the sub-THz-band for each satellite. The available sub-THz wireless frequency from the GBS is split into orthogonal sub-carriers ($\mathcal{C}_U$) in a time-division multiple access (TDMA) fashion, with each sub-carrier having a bandwidth of $W$. This ensures no interference exists between GBS-RIS-RUE \cite{RIS_NO_interference}. Additionally, we assume the GBS and RIS-enabled satellites are deployed by the same service provider.
\vspace{-0.2in} 
\subsection{LEO Satellite Deployment in Orbit \& Geometry}
\vspace{-0.1in}
We consider a circular orbit for each satellite and its three-dimensional (3D) position is obtained as $\boldsymbol{d}_{s}(t)=\big[\left(x_{s}(t), y_{s}(t), z_{s}(t) \right)\big]^{T}, ~\forall s \in \mathcal{S}_m, ~\forall t \in \mathcal{T}.$ Moreover, the next state of satellite $s$ in a certain orbit $m$ of constellation will be: $\boldsymbol{d}^m_{s}(t+1) = \boldsymbol{d}^m_{s}(t) + v_s\Delta_T, ~\forall{t} \in \mathcal{T}, ~\forall{m} \in \mathcal{M}. $
Satellite orientation in the equatorial coordinate system is characterized using three angles, as shown in Fig. \ref{sysmodel}. The inclination angle $i$ represents the intersection angle between the orbital plane and the equator. An inclination angle greater than 90 indicates motion opposite to the Earth's orbit. The angle $\omega$ is between the vernal equinox (normal line exactly above the Equator) and the junctions of the orbital and tropical planes. $\chi$ is the angle formed by the satellite's orientation and the junctions of the orbital and tropical planes. These angles define the location of each satellite $s$. The Cartesian coordinates of each satellite $s$ with the center of the Earth as the origin can be computed using findings from \cite{LEO_coverage_orbit}:
\begin{align} 
    x_s(t) & = (h_s+R_e)\big[\cos{\chi}(t)\cos{\omega}(t) - \sin{\chi}(t)\cos{i}(t)\sin{\omega}(t) \big], ~\forall s \in \mathcal{S}_m, \label{x_coor_sat} \\
    y_s(t) & = (h_s+R_e)\big[\cos{\chi}(t)\sin{\omega}(t) - \sin{\chi}(t)\cos{i}(t)\cos{\omega}(t) \big],  ~\forall s \in \mathcal{S}_m, \label{y_coor_sat}\\
    z_s(t) & = (h_s+R_e)\big[\sin{\chi}(t)\sin{i}(t)\big], ~\forall s \in \mathcal{S}_m,  \label{z_coor_sat}
\end{align}
where $R_e$ is the Earth's radius. We can calculate the distance between two satellites as
     $d_{s,s'} = \sqrt{(R_e+h_s)^2 + (R_e+h_{s'})^2 - 2(R_e+h_s)(R_e+h_{s'})\cos{\delta}},
    \quad \forall {s, {s'} \in \mathcal{S}_m, ~s \neq {s'}},$
where $\delta$ is the angle between two satellites in one orbit as seen from the Earth's center, i.e., $\delta = |\alpha_s -\alpha_{s'}|$, where $\alpha$ is the elevation angle of each satellite $s$. Similarly, the distance between GBS $b$ and the satellite $s$ at each time slot $t$ will be:
    $d_{b,s} = \sqrt{R_e^2\sin^2{\alpha_b} + h_s^2 + 2h_sR_e} - R_e\sin{\alpha_b}, \quad \forall b \in \mathcal{B},~s \in \mathcal{S}_m,~t \in \mathcal{T}.$ 
Moreover, the distance between the satellite $s$ and the RUE $u$ at each time slot $t$ will be:
    $d_{s,u}  = \sqrt{R_e^2\sin^2{\alpha_u} + h_s^2 + 2h_sR_e} - R_e\sin{\alpha_u},~\forall s \in \mathcal{S}_m,~u \in \mathcal{U},~t \in \mathcal{T}. $
Therefore, we can calculate the elevation angle between each GBS $b$ and satellite $s$ at time slot $t$ as:
    $\alpha_{b,s}(t) = \arccos{ \Big( \frac{R_e^2 + d_{b,s}(t) - (R_e+h)^2 }{2R_ed_{b.s}(t)}\Big)  } - \frac{\pi}{2},~
    \forall b \in \mathcal{B},~\forall s \in \mathcal{S}_m, ~\forall t \in \mathcal{T}.$
Similarly, we can find the elevation angle between each satellite $s$ and RUE $u$ at time slot $t$ as:
    $\alpha_{s,u}(t) = \arccos{ \Big( \frac{R_e^2 + d_{s,u}(t) - (R_e+h)^2 }{2R_ed_{s,u}(t)}\Big)  } - \frac{\pi}{2},~
    \forall s \in \mathcal{S}_m,~\forall u \in \mathcal{U}, ~\forall t \in \mathcal{T}.$
The coverage area provided by satellite $s$ depends on the line-of-sight (LoS) propagation distance and the lowest elevation angle $\alpha_{\mathrm{min}}$ at the GBS $b$ and RUE $u$, as mentioned earlier. Moreover, data transmission delay can be configured when utilizing satellite-based RISs for transmitting data packets from the source GBS to destination RUEs is $\tau^s_{b,u} = \frac{d_{b,s} + d_{s,s'} + d_{s,u} }{c}$.
\vspace{-0.2in}
\subsection{LEO Satellite Period in Orbit \& Coverage Region on Earth}
\vspace{-0.1in}
The orbital period, defined as the duration for the mean anomaly to vary by $2\pi$, can be expressed as $\frac{2\pi(h_s+R_e)^{\frac{2}{3}}}{\sqrt{GM_e}}$ \cite{montenbruck2002satellite}, where $G_e$ denotes a Earth-gravitational constant and $M_e$ indicate Earth's mass. To ensure an integer value for the number of periods required for Each rotation, we compute $T_s = \frac{2\pi(h_s+R_e)^{\frac{2}{3}}}{\Delta_T \sqrt{G_eM_e}}$, where $T_s$ denotes how many time slots are necessary for each revolution. All satellites will return to the same positions after $T_s$ time periods, allowing us to analyze a single time frame with $T$ time slots. The position of satellite $s$ in orbit $m$ is obtained from equations (\ref{x_coor_sat} - \ref{z_coor_sat}) by setting $\gamma$ = $\gamma_s$, $i$ = $i_s$, and $\chi$ = $\chi^m_s + 2\pi t/T, t \in \mathcal{T}$. With decreasing elevation angle $\alpha$, the satellite's coverage area reduces \cite{cakaj2014coverage}. Assuming Earth's surface is a perfect globe, the comprised angle $\beta$ from the point with the lowest elevation to the satellite's projection point (angular radius of the coverage circle) is given by $\beta_s = \arccos\Big({\frac{R_e}{R_e+h_s}\cos{\alpha_{\mathrm{min}}}}\Big) - \alpha_{\mathrm{min}}, ~\forall s \in \mathcal{S}_m$. The coverage area of each satellite $s$ is $A_s = 2\pi R^2_e (1 - \cos{\beta}),  ~\forall s \in \mathcal{S}_m$. Consequently, a greater altitude $h_s$ might offer more coverage.
\vspace{-0.2in}
\subsection{Realistic Sub-THz Link Losses}
\vspace{-0.1in}
Three indirect communication links exist GBS-RIS, RIS-RIS, and RIS-RUE. As they share similar features, we use the same channel model. The following subsections discuss the key processes influencing sub-THz signal propagation in the target scenarios. We explore free space path loss, molecule absorption loss, rain, mist, fog, and probable scattering losses, which are the primary factors briefly described as follows. 
\begin{itemize}
    \item The spreading loss accounts for the portion of power emitted by an isotropic antenna operating at frequency $f$, which can be received by an isotropic receiver situated at distance $d$. Assuming spherical propagation, the spreading loss is $ L^{\mathrm{spr}}(f,d) = \frac{4\pi f d}{c^2},$ where $f$ presents the carrier frequency, $c$ depicts the speed of light, and $d$ indicates the distance between the transmitter (Tx) and receiver (Rx). To mitigate the huge channel losses, directional antennas will be utilized at the GBS (Tx) and RUE (Rx). Their directivity gain $D(\alpha, \mu)$, where $\mu$ represents the azimuth angle, in comparison to a perfect isotropic radiator and sensor (detector usually expressed in dBi), ought to be taken into account.
    \item  The sub-THz channel exhibits molecular absorption loss near Earth, which represents the portion of EM energy transformed into kinetic energy within vibrating molecules. This loss is primarily caused by H$_2$O vapors in the air, as they have an absorption spectrum that imparts selectivity to the wireless channel frequency. Therefore, the sub-THz channel between the Tx and Rx is \cite{THz_absorbtion}: $L^{\mathrm{abs}}(f,d) = e^{\kappa_a(f, T_k, p_{\mathrm{atm}})d},$ where $\kappa_a(f)$ denotes the molecular absorption coefficient in $(1/m)$ at frequency $f$, temperature $T_k$, and atmospheric pressure $p_{\mathrm{atm}}$.
    \item  Rain attenuation is a significant challenge for sub-THz satellite communication as signal waves can scatter and be absorbed while traveling through the atmosphere \cite{THz_absorbtion}. ITU-R P.$618$-$13$ defines rain attenuation as $L^{\mathrm{rain}} = \xi_R L_E~$(dB), where $\xi_R$ is the frequency-dependent coefficient (FDC) available in ITU-R P.$838$ \cite{rainattenuation_coeff}, and $L_E$ is the sufficient path length. The FDC of rain $\xi_R$ can be determined as $ \xi_R = \phi_{R} (R_{0.01})^{\mu_R}$ ~ (dB/km), where $R_{0.01}$ represents the rainfall rate (mm/hr), found in ITU-R P.$837$ \cite{rainattenuation_rate}. Additionally, $\phi_R$ and $\mu_R$ are coefficients that depend on frequency, available in ITU-R P.$838$ \cite{rainattenuation_coeff}. Similarly, the cloud loss according to the ITU-R attenuation model \cite{rainattenuation_CLOUD} can be expressed as $L^{\mathrm{cloud}} = \xi_C \chi_C L_E,~$(dB), where $\xi_C$ is the distinct attenuation coefficient, and $\chi_C$ is the density of fluid water $(g/m^3)$ in the cloud or mist (fog).
    \item Sub-THz waves experience additional losses when traveling through the ionospheric plasma, a vast and magnetized plasma ionized by solar radiation. This medium comprises free electrons and ions that impact sub-THz wave propagation, characterized by three metrics: plasma frequency, collision frequency, and depolarization due to Faraday rotation \cite{Thz_channel_modeling_in_space}. 
    
    \textbf{Plasma Frequency Loss:} The frequency of free electrons and ions vibrating in an electric field is given by $f^{\mathrm{plasma}} = \frac{e}{2\pi} \sqrt{\frac{n_e}{\epsilon_0m_e}},$ where $e$ is the charge of an electron, $m_e$ is its mass, $\epsilon_0$ is the permittivity of vacuum, and $n_e$ is the ionospheric electron density.

    \textbf{Collision Frequency Loss:} Collisions occur between charged electrons, ions, and neutral particles, leading to interactions and a collision frequency given as $ f^{\mathrm{col}} = \eta_{\mathrm{ei}} + \eta_{\mathrm{en}} + \eta_{\mathrm{in}},$ where $\eta_{\mathrm{ei}}$, $\eta_{\mathrm{en}}$, and $\eta_{\mathrm{in}}$ represent collision frequencies of electrons with ions, electrons with neutral particles, and ions with neutral particles, respectively. These collision frequencies depend on various factors, including temperature, electron, and neutral particle densities.

    \textbf{Depolarization:} Faraday rotation caused by the ionosphere results in depolarization of sub-THz waves. The depolarization angle over a displacement $S_{\mathrm{TH}}$ is given by $ d\phi_{\mathrm{FR}}/dS_{\mathrm{TH}} = 2.36\times 10^4 \Bar{B}_G f_c^{-2}n_e,$ where $\Bar{B}_G$ is the average intensity of the Earth's magnetic field.

    \textbf{Attenuation Factor:} The attenuation factor, which represents the loss per unit distance along the signal's propagation path, depends on both collision frequency $f^{\mathrm{col}}$ and plasma frequency $f^{\mathrm{plasma}}.$ It can be expressed as $L^{\mathrm{atten}} = \frac{1}{2}k \Bigg(\frac{f^{\mathrm{plasma}}}{f_{c}}\Bigg)^2 \frac{f^{\mathrm{col}}}{f_c} \Bigg( 1 + \frac{1}{2}\bigg( \frac{f^{\mathrm{plasma}}}{f_c} \bigg) \Bigg),$ where $k$ denotes the wave integer, and $f_c$ represent the central frequency. In summary, the total sub-THz wave losses can be defined as:
    \begin{multline}
        \label{total_PL}
        L^{\mathrm{tot}}(f, d) = \frac{(4\pi df)^2 L^{\mathrm{abs}}(f,d) L^{\mathrm{rain}}L^{\mathrm{cloud}}L^{\mathrm{atten}}}{c^2G_{b}G_{u} d_x d_y A^{\alpha} \bigg |\sum_{n_s=1}^{N_r} \frac{\sqrt{F_{n_s}^{\mathrm{tot}}}}{d_{\mathrm{b,s}}d_{\mathrm{s,s'}}d_{\mathrm{s,u}}}\bigg |^{\alpha}},
        ~\forall {b} \in \mathcal{B}, ~\forall {s, {s'} \in \mathcal{S}_m, ~s \neq {s'}}, ~\forall {u} \in \mathcal{U},
    \end{multline}
    where $G_b$ and $G_u$, are the GBS and RUE antenna gains towards directions $\theta_{b}$ and $\theta_{u}$, respectively. The size of a single unit cell along the x-axis and y-axis is represented by $d_x$ and $d_y$, respectively. The considered normalized radiation pattern (NRP) on the obtained signal strength is $ F^{\mathrm{tot}}_{n_s}(\alpha_{n_s}, \mu_{n_s}) = F^{b,s} F^b F^{s,{s'}} F^u F^{s,u},$ where $F^{b,s}(\alpha_{n_s}, \mu_{n_s})$ is the NRP for the elevation angle $\alpha_{n_s}$ and azimuth angle $\mu_{n_s}$ between RIS $n_s$ and GBS $b$, $F^{s,{s'}}$ between RIS $n_{s}$ and RIS $n_{s'}$ where $n_s \neq n_{s'}$, $F^{s,u}$ between RIS $n_{s}$ and RUE $u$, $F^b$ from GBS, and $F^u$ for RUE. Thus, the NRP for given elevation $\alpha$ and azimuth $\mu$ angles are as \cite{satellite_RIS}:
    \begin{equation}
    F(\alpha_{n_s}, \mu_{n_s}) = 
    \begin{cases}
            \cos^3{\alpha}, & \alpha \in \big[ 0, \frac{\pi}{2} \big], ~\mu \in [0, 2\pi] ,\\
 			0,              & \alpha \in \big( 0, \frac{\pi}{2} \big], ~\mu \in [0, 2\pi].
    \end{cases}
    \end{equation}
\end{itemize}
\vspace{-0.2in}
\subsection{Sub-THz Channel Model and Link Analysis}
\vspace{-0.1in}
In the proposed system, each GBS $b$ communicates with $U$ single-antenna RUEs using $K$ antennas in a dedicated Earth region. To extend the coverage range of sub-THz communications, multiple RISs are employed to assist in transmitting $O$ data streams concurrently from the $K$ GBS antennas to the $U$ RUEs. By using RISs, each data stream is beamformed towards a specific RUE after being reflected by the RIS. Multiple RISs are strategically distributed within the satellite constellation, allowing them to adapt their characteristics and capture more energy to enhance transmission distances and overcome path losses. The transmitted signal from GBS $b$ may traverse up to $R_s$ RIS hops before reaching RUE $u$. We define $\boldsymbol{H}_1 \in \mathbb{C}^{(N_1\times K)}$ as the channel matrix between GBS $b$ and the first satellite-based RIS, and $\boldsymbol{H}_{r+1} \in \mathbb{C}^{(N_{r+1}\times N_r)}$ as the channel matrix between RISs $r$ and $r+1$. Consequently, the received signal at RUE $u$ is:
\begin{equation}
    \label{received_signal}
    y_u = \sqrt{\frac{p_{u}}{L^{\mathrm{tot}}}} \boldsymbol{g}^T_u \prod_{r=1,... , R_s} \boldsymbol{\Phi}_r \boldsymbol{H}_r \boldsymbol{x} + z_u, 
\end{equation}
where vector $\boldsymbol{g}_u \in \mathbb{C}^{(N_{R_s}\times 1)}$ captures the channel from the last RIS to RUE $u$, $\boldsymbol{x} \in \mathbb{C}^{(K\times 1)}$ is the communicated vector from GBS $b$ with transmit power $p_u$, and $z_u$ is the additive white Gaussian noise (AWGN) received at RUE $u$ with zero means and $\sigma^2$ variance, i.e., $\mathcal{C}\mathcal{N}(0,\, \sigma^{2})$ variance. The satellite sub-THz channel is dominated by LoS paths, thus, we use Rician fading to simulate the channels $\boldsymbol{H}_r$ by following these works \cite{satellite_RIS, airplance_sat_channel, Thz_channel_modeling_in_space} as $
\boldsymbol{H}_r = \sqrt{\frac{K_H}{K_H+1}}\boldsymbol{\bar{H}}_r + \sqrt{\frac{1}{K_H+1}}\boldsymbol{\tilde{H}}_r,$ where $K_H$ is the Rician factor of $\boldsymbol{H}_r$, $\boldsymbol{\bar{H}}_r$ is the LoS component that remains constant during the channel coherence period, and $\boldsymbol{\tilde{H}}_r$ is the non-LoS (NLoS) component, whose elements are treated as complex Gaussian distributions with zero mean and unit variance, i.e., $\mathcal{C}\mathcal{N}(0,\,1)$. Likewise, the Rician distribution is used to describe the channels $\boldsymbol{g}_u$, with the Rician factor being $K_g$. Let $\boldsymbol{\theta}_r = [\theta_{r1}, \theta_{r2}, ..., \theta_{rn_r}]$ be the reflection coefficient matrix, then the phase shift RIS element responses (i.e., the $r^{\mathrm{th}}$ analog precoding matrix) is $\boldsymbol{\Phi}_r \triangleq  \mathrm{diag}[A_{r1}e^{j\theta_{r1}}, A_{r2}e^{j\theta_{r2}},..., A_{rn_r}e^{j\theta_{rn_r}}] \in  \mathbb{C}^{(N_{r}\times N_r)},$
where $A_{rn}$ is the amplitude of element $n$ in RIS $r$. RISs are considered to have no energy loss, i.e., $A_r = |\boldsymbol{\Phi}_r(n_r,n_r)|^2 = 1$. The effective control of multiple RIS is a significant practical concern, although it lies outside the scope of our research. Relevant studies addressing this matter can be found in reference \cite{RIS_CSI_perfect}. In our work, we assume that both the  BSs and the RIS\footnote{Despite being an idealistic assumption, investigating the performance enhancements achieved by satellite-based RIS in sub-THz communication systems remains a valuable pursuit.} possess perfect knowledge of the CSI, a premise shared by prior work \cite{RIS_CSI_perfect_ref}. 

We also assume each RIS $r$ phase shift resolution is given as $\phi_{n_r} \in [0, 2\pi)~\forall~n_r$. The association matrix for all $U$ RUEs across all $R$ RISs will be given by $\boldsymbol{V} \in\mathbb{C}^{(U \times R)}$. For each RUE $u$, element of the association matrix $\boldsymbol{V}$ will:
\begin{equation}
    v_{b,u}^{s} = 
        \begin{cases}
         1,     & \textrm{if}~\textrm{GBS}~b~\textrm{transmits}~\textrm{to}~\textrm{RUE}~u~\textrm{via}~\textrm{the}~\textrm{s$^\textrm{th}$}~\textrm{satellite},\\
         0,     & \textrm{otherwise}.
        \end{cases}
\end{equation}
The transmitted vector is $\boldsymbol{x} \triangleq \sum_{u=1}^{U} \boldsymbol{w}_u s_u,$ where $\boldsymbol{w}_u \in \mathbb{C}^{(K \times 1)}$ and $s_u \in \mathcal{C}\mathcal{N}(0,1)$, represent, respectively, the beamforming vector and self-dependent RUE symbols, under the assumption of Gaussian signals. The GBS transmit power must satisfy the constraint, i.e., $\varepsilon [|x|^2] = \mathrm{tr}(\boldsymbol{W}^H \boldsymbol{P} \boldsymbol{W}) \leq P^{\mathrm{tot}},$ where $\mathrm{tr}(\boldsymbol{W}^H \boldsymbol{P} \boldsymbol{W})$ is the trace of the square matrix $\boldsymbol{W}$, $ \boldsymbol{W} \triangleq [\boldsymbol{w}_1, \boldsymbol{w}_2,..., \boldsymbol{w}_U]  \in \mathbb{C}^{(K \times U)}$, power variable and $P^{\mathrm{tot}}$ is the GBS's total transmission power. For the received signal given in (\ref{received_signal}), the instantaneous SNR at RUE $u$ at each time slot $t$ is:
\begin{equation}
    \Gamma_{u, t} = \frac{p_{s,u}\bigg| v_{b,u}^{s}\boldsymbol{\mathrm{g}}^T_u(t) \prod_{r=1,... , R_s} \boldsymbol{\Phi}_r \boldsymbol{H}_r(t) \boldsymbol{w}_u\bigg|^2 } {N_o L^{\mathrm{tot}}(f,d)},
\end{equation}
where $N_o$ is noise spectral density. The feasible data rate of RUE $u$ at time slot $t$ is given by $R_{u,t} = B_u \log_2 (1 + \Gamma_{u, t}),$ where $B_u$ is the total available bandwidth for each RUE $u$. Increasing the number of RISs and SNR levels improves the achievable data rate. Next, we will discuss the problem formulation for maximizing the achievable data rate while satisfying the constraints.
\vspace{-0.2in}
\subsection{Problem Formulation}
\label{prob_form}
\vspace{-0.1in}
Our primary goal is to improve the attainable data rate in the sub-THz link propagation attenuation in space through a multi-hop RIS-assisted satellite communication strategy. We aim to optimize the RUE-Satellite association $\boldsymbol{V}$, RIS phase shift $\boldsymbol{\Phi}$, GBS-satellite-RUE data routing $\boldsymbol{\varrho}$, and GBS transmit power $\boldsymbol{P}$ to achieve this. The formulation of the combined optimization problem is as follows:
\begin{maxi!}|s|[2]
		{\substack{\boldsymbol{V},\boldsymbol{\Phi}, \boldsymbol{\varrho}, \boldsymbol{P} }}   
		{ \sum_{u=1}^{U} \sum_{t=1}^{T} R_{u,t} \label{obj1}}{\label{opt:P1}}{\textbf{P1:}}
		\addConstraint{R_{u,t} \geq R^{\textrm{min}},~\forall u \in \mathcal{U}, t \in \mathcal{T}, {\label{c1}}}
		\addConstraint{ v_{b,u}^s \in \left\{0,1 \right\}, ~\forall u \in \mathcal{U}, ~\forall b \in \mathcal{B}, ~\forall s \in \mathcal{S}, t \in \mathcal{T}, {\label{c2}}}
		\addConstraint{ \sum_{b=1}^{B} v_{b,u}^s  = 1, ~\forall u \in \mathcal{U},  ~\forall s \in \mathcal{S}, t \in \mathcal{T}, {\label{c3}}}
		\addConstraint{ 0 \leq \phi_{n_r} \leq 2\pi ,~\forall r \in \mathcal{R}, n \in \mathcal{N}_r, {\label{c5}}}
        \addConstraint{ \sum_{b=1}^{B} \sum_{s=1}^{S} \varrho_{b,u}^s \tau^s_{b,u} \leq \psi_{b,u}^{\textrm{max}}, {\label{c10}} }
        \addConstraint{\mathrm{tr}(\boldsymbol{W}^H\boldsymbol{P}_b\boldsymbol{W}) \leq P^{\mathrm{tot}}_b, ~\forall b \in \mathcal{B}, {\label{c6}} }
		\addConstraint{\lVert \boldsymbol{d}_{s,t} - \boldsymbol{d}_{s',t} \rVert^2_2 \geq D^2_{\mathbf{min}} ,  ~ \forall s\neq s' \in \mathcal{S}, {\label{c7}}}
		\addConstraint{{\lVert \boldsymbol{d}_s(t+1)-\boldsymbol{d}_s(t) \rVert \over{ t^{\textrm{mov}}}} \leq V^\mathbf{max},  ~ \forall s \in \mathcal{S}, t \in \mathcal{T}{\label{c8}} },
		\addConstraint{ \alpha^{\mathrm{LB}} \leq \alpha_{s,t} \leq \alpha^{\mathrm{UB}}, ~ \forall s \in \mathcal{S}, t \in \mathcal{T}{\label{c9}} },
\end{maxi!}
where (\ref{c1}) ensures that the data rate from RUE $u$ at each time slot $t$ will be greater than the minimum threshold that ensures the quality of service (QoS), (\ref{c2}) and (\ref{c3}) show the binary constraints of the RUE-satellite association, which ensure that at each time slot $t$, one RUE $u$ can be associate with at most one satellite $s$, constraint (\ref{c5}) implies that the phase shift values should be between $0$ and $2\pi$, i.e., $\phi_{n_r} \in [0,2\pi)$, constraint ({\ref{c10}}) guarantees that each data packet maximum delay with linked satellites stays below the defined threshold, and constraint ({\ref{c6}}) maintains the total power of the transmit signals from GBS $b$ below a maximum value. The remaining three constraints ensure the satellite constellation design remains intact, i.e., constraint ({\ref{c7}}) ensures that the distance between satellites is not as close as the lowest distance $D_{\textrm{min}}$ to ensure energy efficiency, reduce network interference and maintain optimal elevation angle, constraint ({\ref{c8}}) is a satellite speed constraint that ensures that each satellite completes its cycle within the predefined timeslots, and constraint ({\ref{c9}}) ensures that satellite elevation angle should remain in the feasible region.
\vspace{-0.2in}
\section{Proposed BCD Algorithm Composed of BKMC, MAPPO DRL, and WOA.}
\label{sol_algo}
\begin{figure}[t] 
\centering 
\includegraphics[width=0.5\columnwidth, height=2.3in]{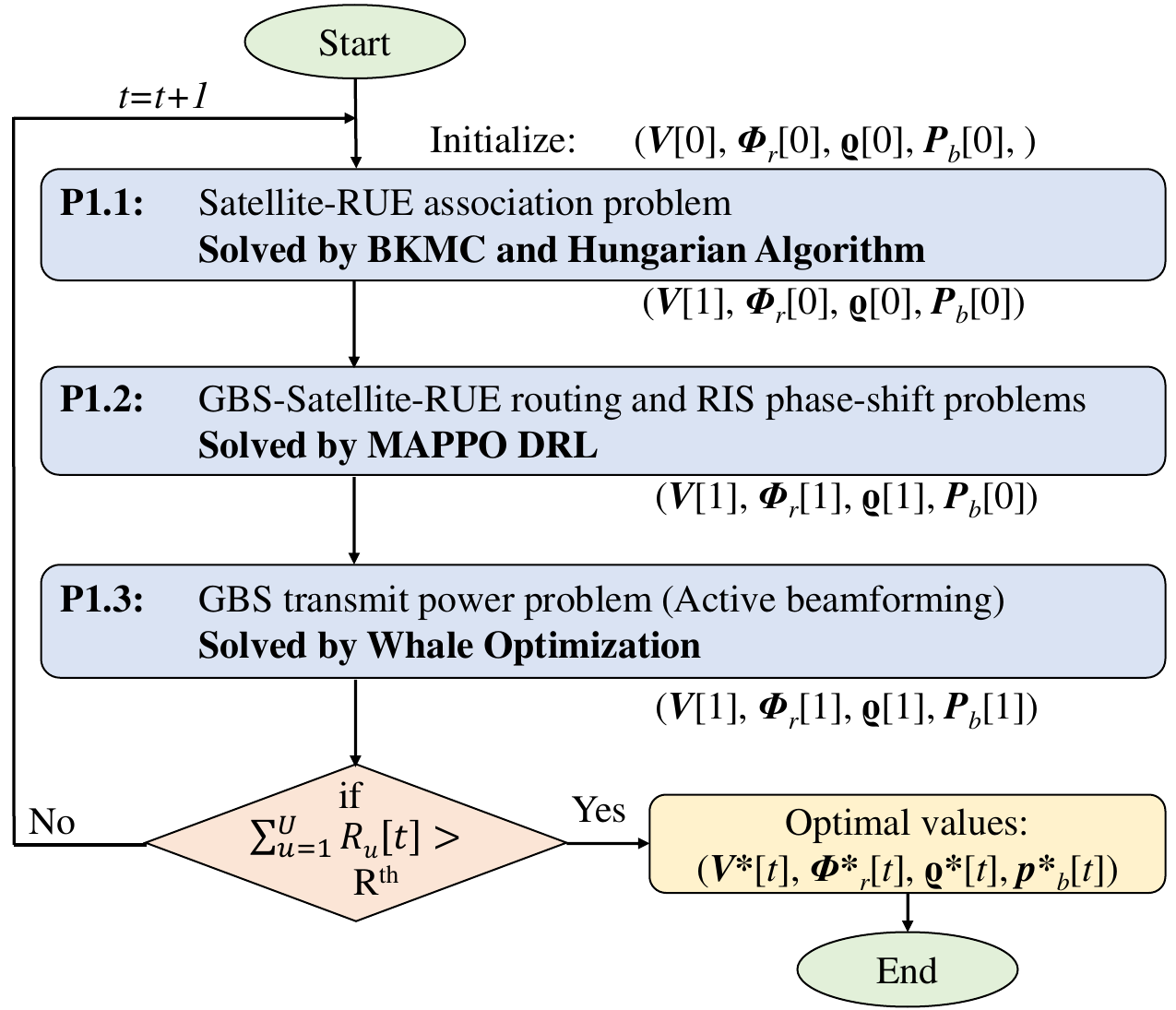}
\caption{Illustration of solution approach in BCD structure.}
\vspace{-0.35in}
\label{BCD}
\end{figure}

\vspace{-0.1in}
We can observe that problem (\ref{opt:P1}) is MINLP due to the decision variables $\{\boldsymbol{V}, \boldsymbol{\Phi}, \boldsymbol{\varrho}, \boldsymbol{p}\}$ that are coupled in the objective function. Due to the combinatorial nature (i.e., NP-hard) of the proposed problem, it is intractable to address this problem in large network settings. Therefore, to address the complexity issue, we adopt the BCD approach owing to its scalable and feasible feature \cite{tseng2001convergence}, where the formulated problem is disintegrated into three tractable subproblems as represented in Fig. \ref{BCD}. Subsequently, the disintegrated subproblems are addressed iteratively using low-complexity algorithms, as described in the subsequent subsections.
\vspace{-0.2in}
\subsection{Balanced K-means Clustering (BKMC) for Satellite-RUE Association} 
\vspace{-0.1in}
We now investigate how to optimize the RUEs' association $\boldsymbol{V}$ for satellites in their coverage regions at each time slot $t$. Therefore given the initial\footnote{Initially, we consider that each decision variable is defined in-bounds and provides the definite value of an objective function.} RIS phase shift $\boldsymbol{\Phi}$, data routing $\boldsymbol{\varrho}$ and GBS transmit power $\boldsymbol{P}$, $\textbf{P1}$ is converted into $\textbf{P1.1}$, which an integer programming. Mathematically, it can be described as follows:
\begin{maxi!}|s|[2]<b>
		{\substack{\boldsymbol{\boldsymbol{V} } }}
		{ \sum_{u=1}^{U} \sum_{t=1}^{T} R_{u,t}(\boldsymbol{V}) \label{obj_P1.1}}{\label{opt:P1_1}}{\textbf{P1.1:}}
		\addConstraint{R_{u,t} \geq R^{\textrm{min}},~\forall u \in \mathcal{U}, t \in \mathcal{T}, {\label{c1_1}}}
		\addConstraint{ v_{b,u}^r \in \left\{0,1 \right\}, ~\forall u \in \mathcal{U}, ~\forall b \in \mathcal{B}, ~\forall r \in \mathcal{R}, {\label{c1_2}}}
		\addConstraint{ \sum_{b=1}^{B} v_{b,u}^r  = 1, ~\forall u \in \mathcal{U},  ~\forall r \in \mathcal{R}, {\label{c1_3}}}
        \addConstraint{\lVert \boldsymbol{d}_{s,t} - \boldsymbol{d}_{s',t} \rVert^2_2 \geq D^2_{\mathbf{min}},   \forall s\neq s' \in \mathcal{S}, {\label{c1_4}}}
		\addConstraint{{\lVert \boldsymbol{d}_s(t+1)-\boldsymbol{d}_s(t) \rVert \over{ t^{\textrm{mov}}}} \leq V^\mathbf{max}},  \forall s \in \mathcal{S}, t \in \mathcal{T},{\label{c1_5}}
		\addConstraint{ \alpha^{\mathrm{LB}} \leq \alpha_{s,t} \leq \alpha^{\mathrm{UB}}, ~ \forall s \in \mathcal{S}, t \in \mathcal{T}{\label{c1_6}} },  
\end{maxi!}
Constraints $(\ref{c1_4})$, $(\ref{c1_5})$, and $(\ref{c1_6})$ are internally linked with the satellite constellations, and ensuring their constraints is crucial while solving the problem. To address the unfairness issue in K-means clustering, we propose the use of BKMC, which calculates the magnitude of each cluster in advance and addresses satellite-RUE association problems \cite{malinen2014balanced, Salman_ICC}. The process is similar to K-means, except for the selection phase. We construct pre-allocated slots based on the sum of RUEs $U$, with each RUE $u$ associated with $U/S$ slots per cluster. If $\lfloor U/S \rfloor = \lceil U/S \rceil = U/S$, all clusters have the identical magnitude. Contrarily, there will be ($U$ mod $S$) clusters with magnitudes of $\lfloor U/S \rfloor$ and $S-$($U$ mod $S$) clusters with magnitudes of $\lceil U/S \rceil$. The selection issue is tackled using the Hungarian method \cite{Hungarian} to discover a selection that minimizes the mean square error (MSE).
\begin{algorithm}[t]
\scriptsize
\caption{\strut BKMC for RUE-Satellite Association} 
	\label{alg:BKMC}
	\begin{algorithmic}[1]
	    \STATE{\textbf{Input:} Initial RUE locations $\{\mathbf{d}_u\}_{u \in \mathcal{U}}$ and satellite locations $\{\mathbf{d}_s\}_{s \in \mathcal{S}}$}. 
		\STATE{\textbf{Compute:} Centroid (satellites) locations $C^0$ of ground clusters nearest to satellite position $\{\mathbf{d}_s\}_{s \in \mathcal{S}}$}.
		\STATE{$t \leftarrow 0$}
		\REPEAT
		\STATE{Compute distances between satellites and RUEs}.
		\STATE{Solve a selection problem by Hungarian algorithm}.
		\STATE{Compute new centroid locations $C^{t+1}$}.
		\UNTIL{the positions of the centroids remain the same}
		\STATE{\textbf{Output:} Optimal satellite-RUE assignments $\boldsymbol{V}^*$.}
	\end{algorithmic}
	\label{Algorithm}
\end{algorithm}
Unlike a standard selection problem with fixed weights, in this case, the weights change dynamically based on the newly found satellites after each K-means iteration. The Hungarian approach is then used to find the pairing with the lowest weight. The update step for selecting new satellite locations is similar to standard K-means centroids, where the new location is determined as the mean of each RUE $u$ location $\boldsymbol{d}_u$ allocated to each cluster $i$, i.e.,
\begin{equation}
C^{(t+1)}_{i} = {1 \over n_{i}} \sum_{\boldsymbol{d}^i_{u} \in C^{(t)}_{i}} \boldsymbol{d}^i_{u},
\end{equation}
where $\boldsymbol{d}^i_{u}$ is the RUE $u$ position in cluster $i$, and $C_i$ is the satellite $s$ location\footnote{Here, the number of clusters $I$ reflects the number of originally deployed satellites $S$.}. The edge weight represents the distance between the satellite $s$ and the RUE $u$, which is updated after each iteration's update step. The BKMC algorithm is described in Algorithm \ref{alg:BKMC}.
\vspace{-0.2in}
\subsection{MAPPO DRL Learning for GBS-Satelite-RUE routing and RIS Phase-shift Control Problem}
\vspace{-0.1in}
By utilizing initial the GBS power $\boldsymbol{P}$ and getting the optimal satellite-RUE association $\boldsymbol{V}^{b*}_{s,u}$ from Algorithm \ref{alg:BKMC}, GBS-satellite-RUE routing $\boldsymbol{\varrho}$ and RIS phase-shift $\boldsymbol{\Phi}$ optimization problem remains MINLP. Mathematically, this problem can be defined as:
\begin{maxi!}|s|[2]
		{\substack{\boldsymbol{\Phi}, \boldsymbol{\varrho}}}
		{ \sum_{u=1}^{U} \sum_{t=1}^{T} R_{u,t}(\boldsymbol{\Phi}, \boldsymbol{\varrho}) \label{obj1.2}}{\label{opt:P1.2}}{\textbf{P1.2:}}
		\addConstraint{R_{u,t} \geq R^{\textrm{min}},~\forall u \in \mathcal{U}, t \in \mathcal{T}, {\label{c1.2_1}}}
		\addConstraint{ 0 \leq \phi_{n_r} \leq 2\pi ,~\forall r \in \mathcal{R}, n \in \mathcal{N}_r, {\label{c1.2.2}}}
        \addConstraint{ \sum_{b=1}^{B} \sum_{s=1}^{S} \varrho_{b,u}^s \tau^s_{b,u} \leq \psi_{b,u}^{\textrm{max}} {\label{c1.2.8}} }.    
\end{maxi!}
Since the objective function in (\ref{opt:P1.2}) accumulates over a long period, we can convert this subproblem into a sequential decision-making problem. To address this modified problem, we will be using MAPPO DRL methods. As shown in Fig. $\ref{Whole_MAPPO}$, each optimization variable $\{\boldsymbol{\Phi}, \boldsymbol{\varrho} \}$ is treated as the PPO agent which can learn simultaneously in the same environment. Furthermore, individual agents derive actions that influence the reward. However, employing the MADRL structure with the same reward is limited in this context, as each agent is subject to different constraints. Hence, we proposed that the GEO satellite offers distinct rewards corresponding to the constraints of each agent. Subsequently, we present the Markov decision process (MDP) of the agents used for learning.
\subsubsection{MDP for Routing Agent}
The factors that vary over time and affect the objective and decision are the present position of all the data packets and the end position of all the data packets. Consequently, the observation $s_{\textrm{RO}}$ of the routing agent at time slot $t$ can be described as below: 
\begin{equation}
     \mathcal{\bar{S}}_{\textrm{RO}}(t) = \{\textrm{current~position~of~all~data~packets,~destination~of~all~the~data~packets}. \}
\end{equation}
After obtaining these observations, the routing agent takes actions following their policy distribution at each time slot $t$, which can be defined as follows:
\begin{equation}
    \mathcal{A}_{\textrm{RO}}(t) = \{ \boldsymbol{\varrho}_{b,u}^s(t) \}, 
\end{equation}
where $\mathcal{A}_{\textrm{RO}}(t)$ and $\varrho_{b,u}^s(t)$ denotes the data routing agent and their actions, i.e., \textit{the optimal direction of packets for routing in each time slot $t$}. After establishing the observations and action spaces, the next step involves defining a reward function that aligns with the optimization problem's objectives while satisfying the relevant constraints. Hence, for the routing agent, the reward function is formulated as follows:
\begin{equation}
     \mathcal{R}_{\textrm{RO}}(t) = \frac{1}{v_{b,u}^{s}\times(d_{b,s} + d_{s,s'} + d_{s,u})},
\end{equation}
where the inverse of the remaining distance ensures less time delay to the data destination.
\subsubsection{MDP for RIS Phase-Shift Agent}
Additionally, the factors that vary over time and affect the objective and decision are the total channel loss and received SNR at RUE $u$. Thus, the observation $s_{\textrm{PS}}$ of the RIS phase-shift agent at time slot $t$ can be described as below: 
\begin{equation}
     \mathcal{\bar{S}}_{\textrm{PS}}(t) = \{L^{\mathrm{tot}}(f, d), \Gamma_{u, t} \}.
\end{equation}
After obtaining these observations, the RIS phase-shift agent takes actions following their policy distribution at each time slot $t$, which can be defined as follows:
\begin{equation}
    \mathcal{A}_{\textrm{PS}}(t) = \{\boldsymbol{\Phi}_s(t)\},
\end{equation}
where $\mathcal{A}_{\textrm{PS}}(t)$ and $\boldsymbol{\Phi}_s(t)$ denotes the RIS phase-shift agent and their actions, i.e., \textit{the selection of optimal phase shift to strengthen the signal in each time slot $t$}. Once the observations and action spaces are constructed, the next crucial step is to define a reward function that aligns with the optimization problem's objectives while satisfying all the associated constraints. Consequently, we can establish the reward function for the RIS phase-shift agent, which can be expressed as:
\begin{equation}
     \mathcal{R}_{\textrm{PS}}(t) = |\boldsymbol{\Phi}\boldsymbol{H}_r|^2,
\end{equation}
where the $R_{\textrm{PS}}(t)$ make the agent to strengthen the channel gain.
\subsubsection{Learning Procedure of MAPPO DRL}
Each agent is essentially an actor-critic module, with the actor in charge of receiving observations and creating action distributions (policies) and the critic in charge of forecasting action value. Each agent is trained using the centralized training and distributed execution (CTDE) strategy \cite{CTDE}. The actor's agent can only view local information and create actions based on those observations. However, the critic may see global information of the other agents to better optimize itself. Let $\pi$ be the each agents' policy, and $V^{\pi}$ signify the state-value function, we have:
\begin{equation}
    V^{\pi}(\mathcal{\bar{S}}(t)) =  \mathbb{E}_{\mathcal{A}(t), \mathcal{\bar{S}}(t+1), \cdots }[R(t)|\mathcal{\bar{S}}(t)], ~~\mathcal{\bar{S}} \in \{\mathcal{\bar{S}}_{\textrm{RO}}, \mathcal{\bar{S}}_{\textrm{PS}} \}.
\end{equation}
The answer to the routing and RIS phase-shift decision problem can be applied to determining the best way to maximize the expected total of discounted rewards for the starting state $\mathcal{\bar{S}}(0)$. As a result, a MAPPO DRL optimization problem is phrased as follows:
\begin{equation}
\begin{aligned}
    \pi^{*}    & =  \arg  \underset{\boldsymbol{\pi}}{\mathrm{max}} ~  \mathbb{E}_{\mathcal{\bar{S}}(0)\sim\rho_s(\mathcal{\bar{S}}(0)) } [V^{\boldsymbol{\pi}} (\mathcal{\bar{S}}(0))], ~~\mathcal{\bar{S}} \in \{\mathcal{\bar{S}}_{\textrm{RO}}, \mathcal{\bar{S}}_{\textrm{PS}} \}.\\
                            & =  \arg \underset{\pi}{\mathrm{max}} ~ \kappa (\pi).  \label{opt_DRL}
\end{aligned}
\end{equation}
Therefore, the goal of the end-to-end satellites data rate maximization problem in (\ref{opt:P1.2}) is represented by (\ref{opt_DRL}) search for an optimal policy to maximize the expected cumulative reward over the period. We consider the PPO strategy to deal with the MADRL problem in (\ref{opt_DRL}) to provide a stable policy learning procedure in dynamic satellite networks. PPO (a policy gradient approach) is less susceptible to hyper-parameters during the training phase than Q-learning-based approaches. In the proposed system, each agent has a policy network (i.e., $\boldsymbol{\theta}^{\varrho^{\textrm{RO}}}$, $\boldsymbol{\theta}^{\varrho^{\textrm{PS}}}$) and a value network (i.e., $\boldsymbol{\sigma}^{\varrho^{\textrm{RO}}}$ and  $\boldsymbol{\sigma}^{\varrho^{\textrm{PS}}}$). Maximizing $\kappa(\pi)$ with provided $\pi^{\mathrm{old}}$ is identical to maximize following:
\begin{equation}
   \mathbb{E}_{\pi}[A^{{\pi}^{\mathrm{old}}}(\mathcal{\bar{S}}(t),\mathcal{A}(t))], ~~\mathcal{\bar{S}} \in \{\mathcal{\bar{S}}_{\textrm{RO}}, \mathcal{\bar{S}}_{\textrm{PS}} \},  ~~\mathcal{A} \in \{\mathcal{A}_{\textrm{RO}}, \mathcal{A}_{\textrm{PS}} \}.
\end{equation}
where $A^{\pi}\big( \mathcal{\bar{S}} (t), \mathcal{A}(t) \big)=\big\{Q^{\pi}( \mathcal{\bar{S}} (t), \mathcal{A}(t)) - V^{\pi}( \mathcal{\bar{S}}(t))\big\}$ identify the advantage function and $Q^{\pi}\big( \mathcal{\bar{S}} (t), \mathcal{A}(t) \big)=\mathbb{E}_{ \mathcal{A}_{(t+1)}, \mathcal{\bar{S}}_{(t+1)}, \cdots }[\mathcal{R}(t)|( \mathcal{\bar{S}} (t), \mathcal{A}(t))]$ is the state-action value function \cite{schulman2017proximal}. Additionally, by approximating $\mathbb{E}_{\pi}\big[A^{{\pi}^{\mathrm{old}}}\big(\mathcal{\bar{S}}(t),\mathcal{A}(t) \big)\big]$ with a clip function, optimization problem (\ref{opt_DRL}) can be converted into:
\begin{equation}
    \underset{\boldsymbol{\theta} = \{\boldsymbol{\theta}^{\varrho^{\textrm{RO}}}, \boldsymbol{\theta}^{\varrho^{\textrm{PS}}}  \}}{\mathrm{max}} \mathbb{E}_{{\pi}^{\mathrm{old}}} \left\{  \mathrm{min}\big[r^{\textrm{ppo}}(\boldsymbol{\theta})A^{{\pi}^{\mathrm{old}}}, ~\mathrm{clip}(r^{\textrm{ppo}}(\boldsymbol{\theta}), 1-\epsilon,1+\epsilon)A^{{\pi}^{\mathrm{old}}} \big] \right\}, \label{PPO}
\end{equation}
where $r^{\textrm{ppo}}(\boldsymbol{\theta}) = \frac{\boldsymbol{\pi}(\mathcal{A}|\mathcal{\Bar{S}};~\boldsymbol{\theta})}{\boldsymbol{\pi}^{\mathrm{old}}(\mathcal{A}|\mathcal{\Bar{S}};~\boldsymbol{\theta}^{\mathrm{old}})}$ is probability ratio. The value of $r^{\textrm{ppo}}(\boldsymbol{\theta})$ is kept inside the range of $[1-\epsilon,1+\epsilon]$ using the clip function, which keeps the policy update to a small range, where $\epsilon$ is a hyperparameter. 
\begin{figure*}[t] 
\centering 
\includegraphics[width=0.8\textwidth, height=3in]{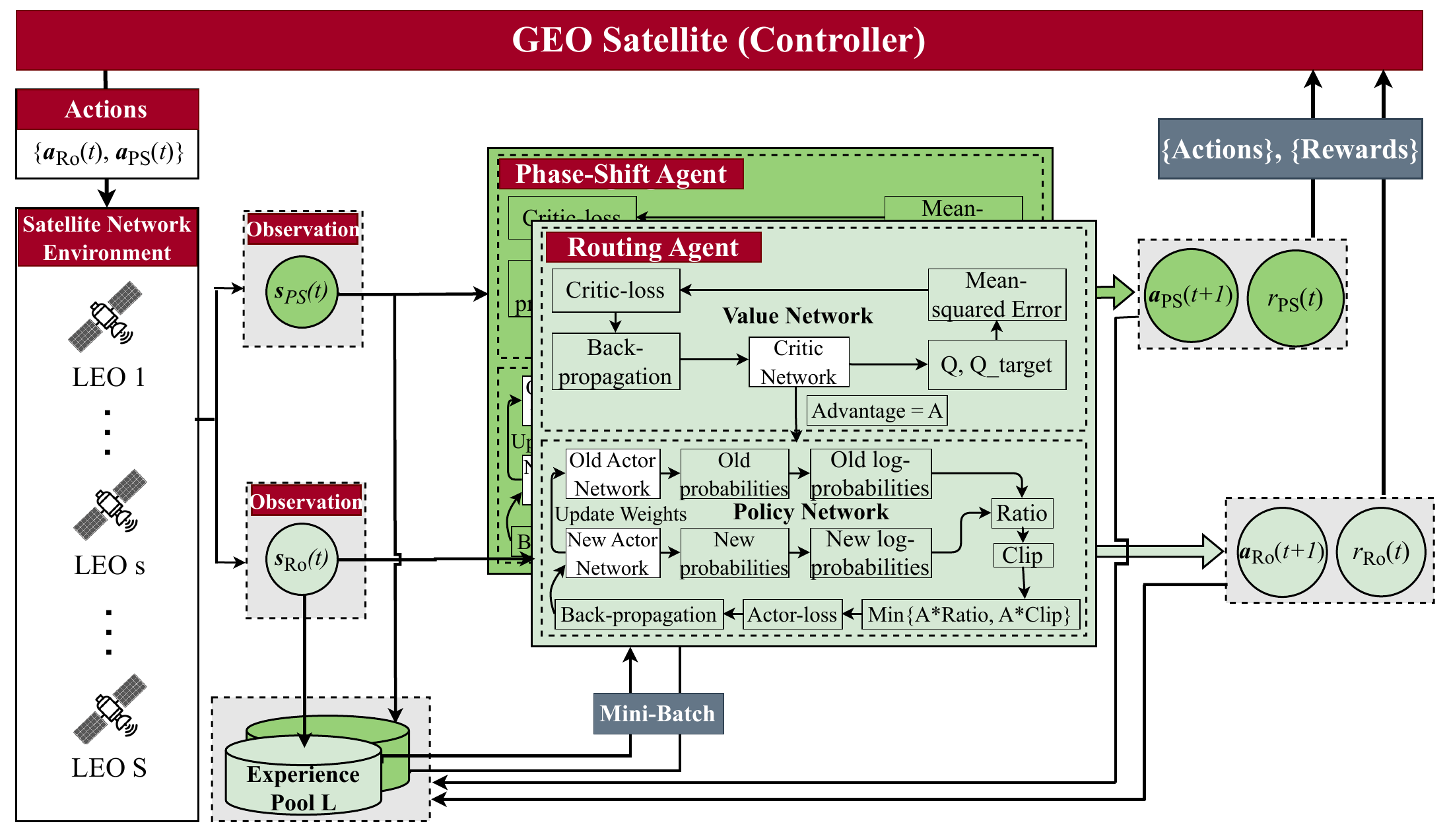}
\caption{Learning structure of multi-agent proximal policy optimization deep reinforcement learning algorithm.}
\label{Whole_MAPPO}
\vspace{-0.4in}
\end{figure*}
The tested action is passed to the critic network when the actor network's parameters are modified, and the critic network estimates the action's estimated value. A strategy gradient approach based on the predicted value \cite{pmlr-v97-iqbal19a} can be used to change the actor. Using preserved samples, the training phase can estimate the expectation of the advantage function $A^{\boldsymbol{\pi}^{\mathrm{old}}}$ in (\ref{PPO}). In other words, the policy is updated using the gradient as follows:
\begin{equation}
    \Delta \boldsymbol{\theta} = \nabla_{\boldsymbol{\theta}} \hat{\mathbb{E}} \Big[ \mathrm{min} \big\{ r^{\textrm{ppo}} (\boldsymbol{\theta}) A^{{\pi}}, \mathrm{clip} (r^{\textrm{ppo}} (\boldsymbol{\theta}), 1-\epsilon, 1+\epsilon) A^{{\pi}} \big\} \Big],  \label{actor_gradient}
\end{equation}
where $A^{\pi}\big( \mathcal{\bar{S}} (t), \mathcal{A}(t) \big) = \hat{Q}\big( \mathcal{\bar{S}} (t), \mathcal{A}(t) \big) - V_{\boldsymbol{\pi}^{\boldsymbol{\mathrm{old}}}}\big( \mathcal{\bar{S}} (t) \big)$ is generalized advantage estimation (GAE) and $\hat{Q}\big( \mathcal{\bar{S}} (t), \mathcal{A}(t) \big)$ can be estimated by using the k-step bootstrapping linear combination i.e., $\hat{Q}\big( \mathcal{\bar{S}} (t), \mathcal{A}(t) \big) = \sum_{k=t}^{\infty} (\Xi \upsilon)^{k-t} \delta(t) + Q_{\boldsymbol{\sigma}^{\mathrm{old}}}\big( \mathcal{\bar{S}} (t), \mathcal{A}(t) \big)$, where $\delta(t) = r(t) + \Xi Q_{\boldsymbol{\sigma}} \big( \mathcal{\bar{S}} (t), \mathcal{A}(t) \big) - Q_{\boldsymbol{\sigma}^{\mathrm{old}}}\big( \mathcal{\bar{S}} (t), \mathcal{A}(t) \big)$ is the temporal difference (TD) \cite{sutton2018reinforcement}. Moreover, each agent seeks to estimate the action-state value function, and thus the value network's loss function is $L^{\mathrm{value}}(t, \sigma) = \big(\hat{Q}\big( \mathcal{\bar{S}} (t), \mathcal{A}(t) \big) - Q_{\sigma^{\mathrm{old}}}\big( \mathcal{\bar{S}} (t), \mathcal{A}(t) \big) \big)^2$. As a result, an optimization problem is devised to train the value network:
\begin{equation}
    \underset{\boldsymbol{\sigma} = \{\boldsymbol{\sigma}^{\varrho^{\textrm{RO}}}, \boldsymbol{\sigma}^{\varrho^{\textrm{PS}}}  \}}{\mathrm{min}} \mathbb{E} \Bigg[  \Big(\hat{Q}\big( \mathcal{\bar{S}} (t), \mathcal{A}(t) \big) - Q_{\boldsymbol{\sigma}^{\mathrm{old}}}\big( \mathcal{\bar{S}} (t), \mathcal{A}(t) \big) \Big)^2 \Bigg]. \label{value_update}
\end{equation}
The gradient descent is utilized to address (\ref{value_update}) which can be define as:
\begin{equation}
    \Delta \boldsymbol{\sigma} = \nabla_{\boldsymbol{\sigma}_s} \hat{\mathbb{E}} \Bigg[  \Big(\hat{Q}\big( \mathcal{\bar{S}} (t), \mathcal{A}(t) \big) - Q_{\boldsymbol{\sigma}^{\mathrm{old}}}\big( \mathcal{\bar{S}} (t), \mathcal{A}(t) \big) \Big)^2 \Bigg],  \label{critic_gradient}
\end{equation}
where parameters $\{\boldsymbol{\theta^{\textrm{RO}}}, \boldsymbol{\theta^{\textrm{PS}}}\}$ and $\{\boldsymbol{\sigma^{\textrm{RO}}}, \boldsymbol{\sigma^{\textrm{PS}}}\}$ can be changed until the loss functions of the policy network and the value network converge, according to (\ref{actor_gradient}) and (\ref{critic_gradient}), respectively. The centralized training approach is used to update the critic network. Each agent's observations and actions are initially dispatched to the TD technique to fine-tune the critic's parameters with the lowest loss function. Fig. \ref{Whole_MAPPO} and Table. \ref{alg:MAPPO} depicts the proposed MAPPO DRL algorithm.
\begin{algorithm}[t]
\scriptsize
	\caption{\strut Learning Process for Multi-Agent Proximal Policy Optimization (MAPPO)} 
	\label{alg:MAPPO}
	\begin{algorithmic}[1]
	    \STATE{\textbf{Initialize:} the initial network $\pi^{\varrho^{\textrm{RO}}}_{0}$ and $\pi^{\Phi^{\textrm{PS}}}_{0}$ for agent of routing and phase shift}
	    \STATE{\textbf{Obtain:} optimal BS-satellite-RUE association from \textbf{P1.1} solution, i.e., $\boldsymbol{V}^{*}$}
		\FOR{episode$=1,2,...,E$}
		\STATE{Initialize randomly each RUE's and GBS position.}
		\FOR{time slot$=1,2,...,N$}
		\STATE{Upload observation $\mathcal{\Bar{S}}_{\textrm{RO}}(t)$ and $\mathcal{\Bar{S}}_{\textrm{PS}}(t)$ of each LEO satellite $s$ to GEO satellite}
		\STATE{Run policy $\mathcal{A}^{\varrho^{\mathrm{RO}}}_{v}\sim\pi^{\varrho^{\mathrm{RO}}}_{\boldsymbol{\theta}_{\mathrm{old}}}$ and $\mathcal{A}^{\Phi^{\mathrm{PS}}}_{v}\sim\pi^{\Phi^{\textrm{PS}}}_{\boldsymbol{\theta}_{\textrm{old}}}$}
		\STATE{Compute each reward $\mathcal{R}^{\varrho^{\textrm{RO}}}_{v}$ and $\mathcal{R}^{\Phi^{\textrm{PS}}}_{s}$ for satellite $s$}
		\STATE{Save $\big(\mathcal{\Bar{S}}_{\textrm{RO}}(t),\mathcal{A}_{\textrm{RO}}(t),\mathcal{R}_{\textrm{RO}}(t),\mathcal{\Bar{S}}_{\textrm{RO}}(t+1) \big)$ and $\big(\mathcal{\Bar{S}}_{\textrm{PS}}(t),\mathcal{A}_{\textrm{PS}}(t),\mathcal{R}_{\textrm{PS}}(t),\mathcal{\Bar{S}}_{\textrm{PS}}(t+1)\big)$ in memory of routing and RIS phase-shift agent respectively}
		\ENDFOR
		\STATE{Compute advantage estimates $\left\langle \hat{A}^{\varrho^{\textrm{RO}}}_{1},...,\hat{A}^{\varrho^{\textrm{RO}}}_{N} \right\rangle$, $\left\langle \hat{A}^{\Phi^{\textrm{PS}}}_{1},...,\hat{A}^{\Phi^{\textrm{PS}}}_{N} \right\rangle$}
		\STATE{Optimize PPO objective (\ref{PPO}) wrt $\boldsymbol{\theta}^{\varrho^{\textrm{RO}}}$ and $\boldsymbol{\theta}^{\Phi^{\textrm{PS}}}$ with minibatch from memory}
		\STATE{$\boldsymbol{\theta}^{\varrho^{\textrm{RO}}}_{\textrm{old}} \leftarrow \boldsymbol{\theta}^{\varrho^{\textrm{RO}}}$, $\boldsymbol{\theta}^{\Phi^{\textrm{PS}}}_{\textrm{old}} \leftarrow \boldsymbol{\theta}^{\Phi^{\textrm{PS}}}$}
		\ENDFOR
		\STATE{\textbf{Output:} Optimal networks $\pi^{\varrho^{\textrm{RO}}}_{\boldsymbol{\theta^*}_{\textrm{opt}}}$ and $\pi^{\Phi^{\textrm{PS}}}_{\boldsymbol{\theta^*}_{\textrm{opt}}}$}
	\end{algorithmic}
\end{algorithm}
\vspace{-0.2in}
\subsection{GBS Transmit Power Optimization based on WOA}
\vspace{-0.1in}
In this part, we will deal with how to optimize the GBS transmit power (i.e., active beamforming) $\boldsymbol{P}$ at each time slot $t$. Thus, given the optimal satellite-RUE association $\boldsymbol{V}^*$, RIS phase-shift $\boldsymbol{\Phi}^*$, and data routing $\boldsymbol{\varrho}^*$, the remaining problem is converted into $\textbf{P1.3}$. Mathematically, it can be described as follows:
\begin{maxi!}|s|[2]<b>
		{\substack{\boldsymbol{\boldsymbol{P}}}}
		{ \sum_{u=1}^{U} \sum_{t=1}^{T} R_{u,t}(\boldsymbol{P}) \label{obj1.1}}{\label{opt:P1_1}}{\textbf{P1.3:}}
		\addConstraint{R_{u,t} \geq R^{\textrm{min}},~\forall u \in \mathcal{U}, t \in \mathcal{T}, {\label{c2_1}}}
		\addConstraint{ \sum_{s=1}^{S} P_{b,s} \leq P^{\textrm{max}}, ~\forall b \in \mathcal{B}, {\label{c2_2}}}
		\addConstraint{\mathrm{tr}(\boldsymbol{W}^H \boldsymbol{P}_s\boldsymbol{W}) \leq P^{\mathrm{tot}}.{\label{c2_3}} }
\end{maxi!}
We use a meta-heuristic algorithm based on WOA to address the non-convex problem in (\ref{opt:P1_1}), inspired by whale prey hunting behavior. WOA has several benefits \cite{mirjalili2016whale}: first, it simplifies gradient computation and step size updating compared to gradient-based algorithms. Second, it is not affected by initial feasible solutions, reducing convergence dependence on the starting point. Third, it balances exploration and exploitation effectively. Fourth, it improves the likelihood of avoiding local optima, making it suitable for large-scale network problems. The WOA is composed of two stages, i.e., exploitation and exploration that help us reach near-optimal solutions. Before applying WOA, we briefly explain the preliminaries of this algorithm in the following subsections \cite{Sone_WOA}.
\subsubsection{\textbf{ Exploitation Phase}} The WOA exploitation phase contains two main methods: encircling target and spiral bubble-net attack. \\
\textbf{Encircling Prey:} Humpback whales may detect the presence of a target (e.g., krill) and encircle it. The WOA algorithm considers that the present most suitable tracking agent is the target prey (optimum or near to optimum), and other whales (tracking agents) adjust their location about the most suitable tracking agent during iterations using the subsequent equations:
\begin{equation}
    \overrightarrow{D} = \big|\overrightarrow{C} \cdot \overrightarrow{X}^*(t) - \overrightarrow{X}(t) \big|, \label{Exploit_D}
\end{equation}
\vspace{-.35in}
\begin{equation}
    \overrightarrow{X}(t+1) = \lfloor \overrightarrow{X}^*(t) -  \overrightarrow{A} \cdot \overrightarrow{D} \rfloor, \label{Exploit_X_UP}
\end{equation}
where $t$ is the present iteration, $\overrightarrow{X}^*(t)$ is the most useful tracking agent's location, $\overrightarrow{X}(t)$ is the position vector, $|\cdot|$ is the absolute value, $\lfloor \cdot \rfloor$ is the space function in discrete form for the channel, and $\cdot$ is element-wise multiplication. Noting that $\overrightarrow{X}(t)$ should be modified in each iteration if a better solution exists. The coefficient vectors $\overrightarrow{A}$ and $\overrightarrow{C}$ are computed as follows:
\begin{equation}
    \overrightarrow{A} =  2\overrightarrow{a} \cdot  \overrightarrow{r} - \overrightarrow{a}, \label{A_value}
\end{equation}
\vspace{-.35in}
\begin{equation}
    \overrightarrow{C} =  2 \cdot \overrightarrow{r}, \label{C_value}
\end{equation}
where control parameter vector $\overrightarrow{a}$ decreases linearly from $2$ to $0$ across iterations and $\overrightarrow{r}$ is a arbitrary vector in the range $[0, 1]$, both in exploration and exploitation. The primary goal of (\ref{A_value}) and (\ref{C_value}) is to strike a balance between exploration and exploitation. The exploration is conducted when $A \geq 1$, and the exploitation is when $A < 1$. The maximum number of iterations is represented by $I_{\mathrm{max}}$, and the managing parameter $\overrightarrow{a}$ can be revised as $\overrightarrow{a} = 2(1-t/I_{\mathrm{max}})$. 
\textbf{Bubble-Net Attacking Method:} The decreasing encircling and spiral revising position processes are employed in tandem to represent the whale's bubble-net attacking method. The shrinking encircling process is performed by placing the coefficient vector $\overrightarrow{A}$ in $[-1, 1]$ while progressively decreasing the value of $\overrightarrow{a}$ over repetitions. As a result, the new location will be positioned between the agent's present location and the position of the most suitable tracking agent. The spiral equation between the position of the target and the whale can be utilized to simulate the helix-shaped motion of whales as follows:
\begin{equation}
    \overrightarrow{D'} = \big| \overrightarrow{X}^*(t) - \overrightarrow{X}(t) \big|,
    \label{Exploit_D'}
\end{equation}
\vspace{-.35in}
\begin{equation}
    \overrightarrow{X}(t+1) = \lfloor \overrightarrow{D'} \cdot \exp^{bl} \cdot \cos{(2\pi l)} + \overrightarrow{X}^*(t)  \rfloor, \label{Exploit_X'_up}
\end{equation}
where $\overrightarrow{D'}$ is the distance between the target and the present tracking agent, $b$ is the logarithmic spiral shape's constant, and $l$ is a random value in the range $[-1, 1]$. The shrinking encircling method and the spiral approach are utilized together because whales swim around their prey in a shrinking circle while also moving along a spiral-shaped path. To describe this behavior, each method is considered to be done with a chance of $50\%$ as follows:
\begin{equation}
\overrightarrow{X}(t+1) =
 		\begin{cases}
 			\overrightarrow{X}^*(t) - \overrightarrow{A} \cdot \overrightarrow{D}, & \text{if~} p < 0.5,\\
 			\overrightarrow{D'} \cdot \exp^{bl} \cdot \cos{(2\pi l)} + \overrightarrow{X}^*(t),  & \text{if~} p \geq 0.5,\\
 		\end{cases}
\end{equation}
where $p$ is an arbitrary value in the range $[0, 1]$ denotes the probability of selecting one of two methods, i.e., WOA selects the shrinking encircling method if $p<0.5$ and WOA selects the spiral movement method if $p\geq0.5$.
\subsubsection{\textbf{Exploration Phase}} WOA's exploration phase contains the method of hunting for prey. This phase is required to keep the solution from being stuck at the local optimum and failing to obtain the global optimum.\\
\textbf{Search for Prey:} The prey search can be approached using a methodology similar to the shrinking-encircling process. The location $\overrightarrow{X}^*(t)$ of the most suitable tracking agent is now substituted by the position $\overrightarrow{X}_{\mathrm{rand}}(t)$ of a whale chosen at arbitrary from the present inhabitants, and the coefficient vector $\overrightarrow{A}$ with $|\overrightarrow{A}|>1$ is used. E.g., whales are driven to migrate away from an arbitrary whale, allowing the WOA algorithm to expand the tracking field and conduct a global search. The following is the mathematical representative for the target hunt:
\begin{equation}
    \overrightarrow{D} = \big|\overrightarrow{C} \cdot \overrightarrow{X}_{\mathrm{rand}}(t) - \overrightarrow{X}(t) \big|, \label{Explore_D}
\end{equation}
\vspace{-.35in}
\begin{equation}
    \overrightarrow{X}(t+1) = \lfloor \overrightarrow{X}_{\mathrm{rand}}(t) -  \overrightarrow{A} \cdot \overrightarrow{D} \rfloor, \label{Explore_X_up}
\end{equation}
\vspace{-0.4in}
\subsubsection{\textbf{Fitness function for the constraint}}
Because the WOA algorithm is primarily designed for unconstrained optimization, we use the penalty approach to cope with the problem's date rate fairness constraint in (\ref{c2_1}) \cite{WOS_penalty}. In (\ref{opt:P1_1}), RUEs are assumed as tracking agents (humpback whales), and the GBS power control $\boldsymbol{P}$ denoted as the position of the tracking agents $\boldsymbol{X}$. The power control $\boldsymbol{P}(t)$ (corresponding to $\boldsymbol{X}(t)$) can be updated at iteration $t$ by pursuing either the shrinking encircling mechanism, the spiral updating position, or the tracking for the target. Thus we designed the fitness function used to select the optimal tracking agent as follows:
\begin{equation}
    \mathrm{Fitness}(\boldsymbol{P}) = -\sum_{s=1}^{S} \sum_{t=1}^{T} \Big \{ R_{s,t}(\boldsymbol{P}) + \mu F_{s,t}(f_{s,t}(\boldsymbol{P})) f^2_{s,t}(\boldsymbol{P})   \Big \} \label{fitness}
\end{equation}
where $f_{s,t}(\boldsymbol{P})~=~R^{\mathrm{min}} - R_{s,t}(\boldsymbol{P})$ and $\mu~=~10^{14}$. It is worth noting that putting a negative symbol before the objective function is a method for converting a maximizing problem to a minimization problem. The inequality function $f_{s,t}(\boldsymbol{P})$ is redefined as $f_{s,t}(\boldsymbol{P})~=~R_{s,t}(\boldsymbol{P})-R^{\mathrm{min}}$ and the index function $F_{s,t}(f_{s,t}(\boldsymbol{P}))~=~0$ if $f_{s,t}(\boldsymbol{P})~\geq~0$ and $F_{s,t}(f_{s,t}(\boldsymbol{P}))~=~1$ if $f_{s,t}(\boldsymbol{P})~<~0$. The method for optimal power control $\boldsymbol{P}^*$ calculation with WOA is shown in Algorithm \ref{alg:WOA}.
\begin{algorithm}[t]
\scriptsize
	\caption{\strut Whale Optimization Algorithm for GBS Power Control Optimization} 
	\label{alg:WOA}
	\begin{algorithmic}[1]
	    \STATE{\textbf{Input:} the current power control $\boldsymbol{p}$, optimal Satellite-RUE association $\boldsymbol{V}^*$, data routing $\boldsymbol{\varrho}^*$ and RIS phase-shift $\boldsymbol{\phi}^*$} 
		\STATE{\textbf{Initialize:} the agent inhabitants $\boldsymbol{p}_{s}$, $s \in \left\{ 1,...,S \right\}$, iteration $t=1$, maximum number of iterations $T_{\mathsf{max}}$.}
		\STATE{Compute the fitness of the tracking agents $\boldsymbol{P}_{s}$ by \eqref{fitness} and set the most suitable tracking agent $\overrightarrow{\boldsymbol{P}}^*(0)$}.
		\WHILE{$t < T_{\mathsf{max}}$}
		\FOR{$\mathcal{S}=1,2,...,S$ (the number of agents)}
		\STATE{Update $a,A,C,l$ and $P$.}
		\IF{$p < 0.5$}
		\IF{$\left| A \right| < 1$}
		\STATE{Update $\Vec{D}$ by \eqref{Exploit_D} and $\Vec{\boldsymbol{P}}$ by \eqref{Exploit_X_UP}}
		\ELSE
		\STATE{Select a arbitrary $\overrightarrow{\boldsymbol{P}}_{\mathrm{rand}}$ and update $\Vec{D}$ by \eqref{Explore_D}}
		\STATE{Update the position $\Vec{\boldsymbol{P}}$ by \eqref{Explore_X_up}}
		\ENDIF
		\STATE{Update $\Vec{D}$ by \eqref{Exploit_D'} and $\Vec{\boldsymbol{P}}$ by \eqref{Exploit_X'_up}}
		\ENDIF
		\ENDFOR
		\STATE{Compute the fitness of each search agent by \eqref{fitness}.}
		\STATE{Update position $\boldsymbol{P}^{*}(t)$ of the most suitable tracking agent.}
		\STATE{$t \leftarrow t+1$}
		\ENDWHILE
		\STATE{\textbf{Output:} The optimal GBS power control $\boldsymbol{P}^*$.}
	\end{algorithmic}
\end{algorithm}
\vspace{-0.2in}
\section{Simulation Settings and Results Discussion}
\label{simul}
\vspace{-0.1in}
By taking into consideration, the 3D coordinates of the satellites $\boldsymbol{d}_s$, each satellite $s$ orbits at a fixed altitude of $h_s$ = $500$km with a constant speed $v_s$ following our proposed deployment scheme. In each initial orbital plane, the satellite is considered as an equally spaced interval and orbiting circularly. The RUEs follow the HPPP to arrive at the AoI. The rest of the main simulation parameters are given in Table. \ref{sim_tab}. The baselines are defined as \textbf{MA-PPO:} The MAPPO algorithm is introduced, where each optimization variable is considered as a PPO agent, responsible for learning its variable optimization. \textbf{MA-TRPO:} The algorithm treats each agent as a trust region policy optimization (TRPO)-based DRL learning agent \cite{trpo}, which seems most related to the proposed PPO algorithm. \textbf{MA-A2C} The algorithm under consideration treats each agent as an Advantage Actor-Critic (A2C)-based DRL learning agent. \cite{baslines_MAA2C}. \textbf{Central:} The algorithm, which considers all decision variables as a single agent, coordinates centrally to optimize them. \textbf{NonRIS:} Instead of using a satellite-based RIS, we only employed a basic satellite to compare the two scenarios \cite{baslines_relay_satellite}. \textbf{Fairness$_{\textrm{p}}$:} This method considers a fair power allocation from each GBS. However, it optimizes factors other than power allocation using the proposed approach \cite{baslines_fairness}.
\setlength{\arrayrulewidth}{0.08mm}
\setlength{\tabcolsep}{1pt}
\renewcommand{\arraystretch}{0.6}
\begin{table}[t]
\centering
\caption{Simulation Parameters}
\label{sim_tab}
\scalebox{1}{
\begin{tabular}{|l|l||l|l|}
\hline
    \textbf{Parameter}& \textbf{Value} & \textbf{Parameter} & \textbf{Value} \\ \hline \hline
    Electron charge & $e~$=$~1.6021\times10^{-19}$~C &  Satellite altitude & $h_s~=~500~$km \\ \hline
    Electron mass & $m_e~$=$~9.109\times 10^{-31}$~kg & Satellite per orbital plane & $S_m~=~22~$ \\ \hline
    Vacuum permittivity & $\epsilon_0~$=$~8.854\times 10^{-12}$~F/m & Minimum elevation angle $\alpha_{\mathrm{min}}$ & $~12^{\circ}$ \\ \hline
    Speed of light & $c~$=$~299,792,458$~m/s & Length of orbital plane & $l_s~=~43486~$km \\ \hline
    Radius of Earth & $R_e~$=$~6,378,100$~m   & Initial distance between satellite & $~1976~$km \\ \hline
    Radius of orbital plane & $R_s~$=$~6,878,100$~m   & Time slot size & $\Delta_T~=~10$ \\ \hline
    Carrier frequency & $f_c~$=$~0.1$~THz & Number of time slot for a episode & $T_E~=~513$ \\ \hline
    Average temperature & $T_{\mathrm{tem}}~$=$~1000$~K & Learning rate & $lr~=~0.0003$ \\ \hline
    Mass of ions & $m_{\mathrm{ion}}~$=$~1.67493\times 10^{-27}$~kg &  Epoch, Batch size, Discount factor & $~3$, $~16$, $\gamma~=~0.95$  \\ \hline
    Neutral particle density & $n_n~$=$~3\times 10^{5}$/cm$^3$ & Training iterations, No. of iterations/update & $~500000$, $~16$ \\ \hline
    Earth magnetic field intensity & $\Bar{B}_G~$=$~25-65~$$\mu T$ & Actor, Critic Network & $[128,128]$, $[16,16]$ \\ \hline
\end{tabular}}
\vspace{-0.25in}
\end{table}
\begin{figure}[t]
\centering
\begin{minipage}{0.5\textwidth}
    \centering
    \includegraphics[width=\columnwidth, height=2.2in]{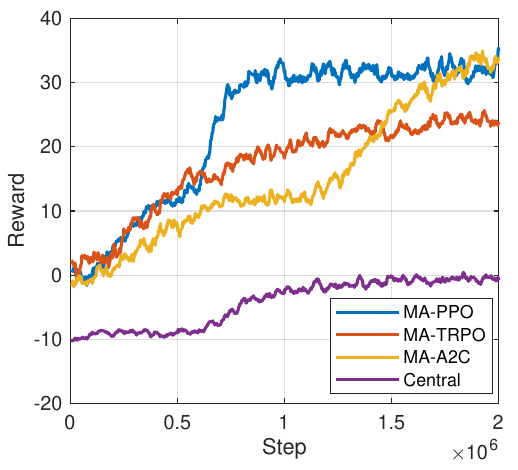}
    \caption{RIS phase-shift agent learning convergence.}
    \label{RIS_agent_convergence}
\end{minipage}%
\begin{minipage}{0.5\textwidth}
    \centering
    \includegraphics[width=\columnwidth, height=2.2in]{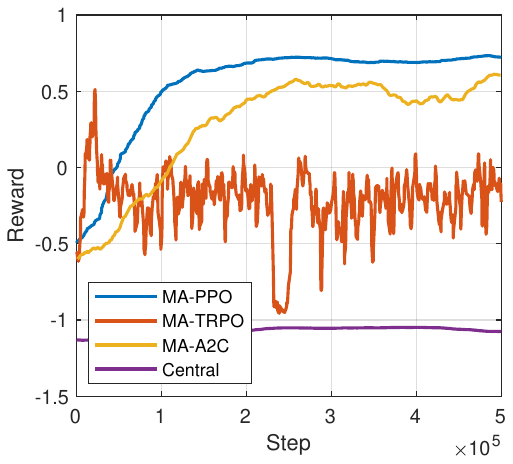}
    \caption{Routing agent learning convergence.}
    \label{Route_agent_convergence}
\end{minipage}
\vspace{-0.4in}
\end{figure}

Fig. \ref{RIS_agent_convergence} depicts the convergence behavior of the phase-shift agent in the satellite-based RIS system concerning the time step. A comparative analysis was performed, involving the MA-PPO-based agent and three baseline algorithms, i.e., MA-TRPO, MA-A2C, and Central agent. The results indicate that the proposed MA-PPO-based algorithm exhibits rapid convergence, achieving higher rewards compared to MA-TRPO by $40\%$, MA-A2C by $8.69\%$, and Central agent, which performed worst (as commonly considered in current literature for solving the overall joint optimization problem). These findings support the suitability and effectiveness of the MA-PPO-based algorithm for the proposed RIS-based satellite networks. Moreover, it is noteworthy that the MA-PPO algorithm outperforms the baseline approaches in terms of stability.

Likewise, Fig. \ref{RIS_agent_convergence} illustrates the convergence behavior of the routing agent in the satellite-based Reconfigurable Intelligent Surface (RIS) system concerning the time step. A comparative analysis was carried out, involving the MA-PPO-based agent and three other baseline algorithms: MA-TRPO, MA-A2C, and Central agent. The results demonstrate that the proposed MA-PPO-based algorithm exhibits rapid convergence, yielding higher rewards compared to the underperforming MA-TRPO, a modest improvement of $22.22\%$ over MA-A2C, and significantly outperforming the central agent, which also showed the weakest performance. These findings provide evidence of the suitability and effectiveness of the MA-PPO-based algorithm for the RIS-based satellite networks proposed in this study. Moreover, it is noteworthy that the MA-PPO algorithm consistently outperforms the other baseline approaches in terms of stability.
\begin{figure}[t]
\centering
\begin{minipage}{0.5\textwidth}
                 \centering
                \includegraphics[width=\linewidth, height=2.2in]{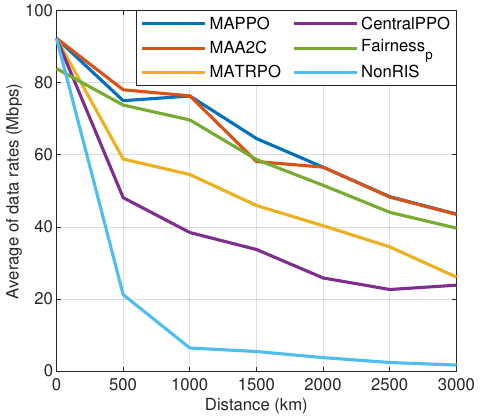}
                \caption{Average of date rates according to distance.}
                \label{performance_rate}
\end{minipage}%
\begin{minipage}{0.5\textwidth}
                \centering
                \includegraphics[width=\linewidth, height=2.2in]{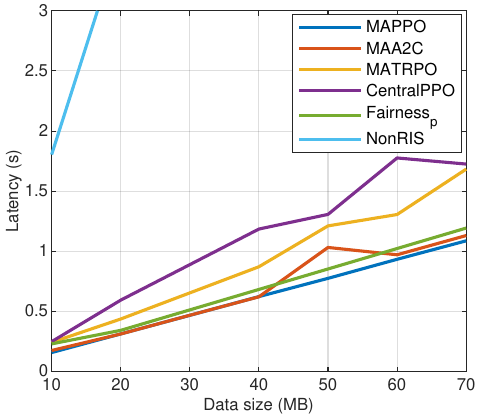}
                \caption{Latency according to data size.}
                \label{performance_latency}
\end{minipage}
\vspace{-0.4in}
\end{figure}

Fig. \ref{performance_rate} presents the important comparative analysis of the proposed MAPPO algorithm with several baselines, namely MAA2C, MATRPO, CentralPPO, Fairness$_{\textrm{p}}$, and NonRIS. The evaluation involves assessing the average data rate as a function of the varying distance of data packets from the source satellite to the destination AoI. The results reveal a consistent trend where the average data rate decreases with increasing distance between the source satellite and the destination AoI. Notably, the proposed MAPPO algorithm consistently outperforms the other baseline approaches in terms of overall performance. Despite MAA2C demonstrating a competitive performance with a close average data rate to MAPPO, it exhibits instability in the results, making the proposed MAPPO algorithm a more reliable and favorable choice. Additionally, the NonRIS case performs the worst in terms of average data rate, primarily due to substantial signal losses in the space environment, which RIS technology can effectively address. This finding further strengthens the proposition of utilizing RIS-based satellites for enabling efficient worldwide communication in future 6G networks, particularly in mitigating signal losses.

Likewise, Fig. \ref{performance_latency} presents results about one of the primary contributions of this research. The graph illustrates the latency (data delay) achieved by the proposed MAPPO algorithm concerning the data packet size, as compared to the performance of the baseline methods. The findings demonstrate that the proposed MAPPO algorithm consistently outperforms the other baselines in terms of both latency and result stability. While the MAA2C approach showed competitive results, the overall performance of MAPPO remained superior due to its higher level of result stability. Furthermore, the NonRIS scenario exhibited the poorest performance, primarily attributed to substantial signal losses. This observation reinforces the viability of our proposal to utilize RIS in satellite-based systems as the optimal approach for future 6G networks. The results support the efficacy of the MAPPO algorithm in addressing latency challenges and highlight the potential benefits of integrating RIS technology for enhancing the performance and reliability of future 6G networks.
\begin{figure}[t]
\centering
\begin{minipage}{0.5\textwidth}
                 \centering
                 \captionsetup{justification=centering,singlelinecheck=false}
                \includegraphics[width=\linewidth, height=2.2in]{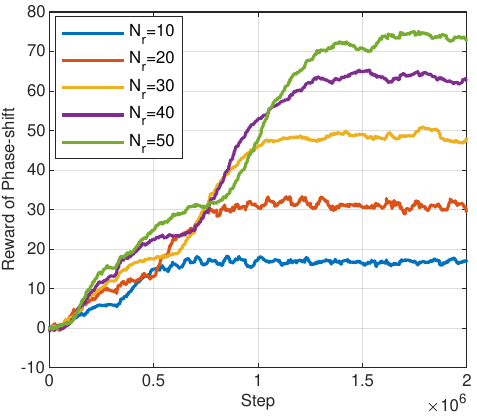}
                \caption{Learning convergence via RIS elements.}
                \label{performance_ris_conv}
\end{minipage}%
\begin{minipage}{0.5\textwidth}
                \centering
                \captionsetup{justification=centering,singlelinecheck=false}
                \includegraphics[width=\linewidth, height=2.2in]{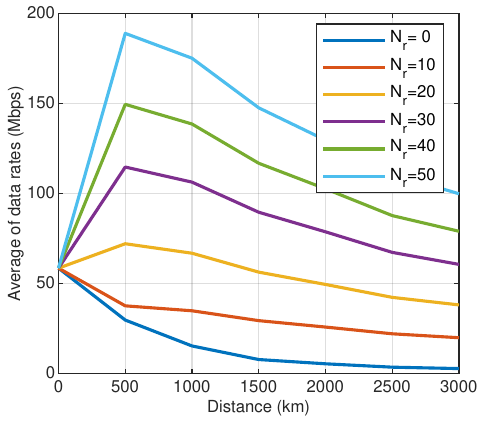}
                \caption{Data rate vs distance.}
                \label{performance_datarate}
    \end{minipage}
    \vspace{-0.4in}
\end{figure}

In Fig. \ref{performance_ris_conv}, we explore the impact of RISs and their number of elements on the rewards achieved by the RIS-phase shift agents over iterations. The graph shows that increasing the number of RIS elements leads to higher rewards for the RIS phase-shift agent. This positive correlation suggests that the size of the RIS influences the agent's reward acquisition performance. Additionally, we observe that each configuration reaches a point of convergence, resulting in stable outcomes. This implies the existence of an optimal RIS size or configuration beyond which further increases in elements may not significantly enhance the agent's rewards, leading to stable performance levels. These insights hold relevance for the design and optimization of RIS-based communication systems. Further analysis of the factors contributing to these trends will enhance our understanding of the intricate interplay between RIS size, agent rewards, and overall system performance.

Fig. \ref{performance_datarate} displays the outcomes of average data rates concerning the distance while varying the number of RIS elements. The graph reveals a noteworthy trend wherein the average data rates increase as the number of RIS elements grows, but only beyond a certain threshold. When the number of RIS elements is equal to or exceeds 20 ($\textrm{N}_r \geq 20$), the average data rates experience an upsurge. However, for cases where the number of RIS elements is insufficient, the data rates decrease due to the challenges associated with long-distance communication and signal losses. Moreover, a subsequent analysis indicates that initially, with 20 or more RIS elements, the average data rate increases notably. Nevertheless, as the distances between communicating nodes extend beyond 500 km, the performance begins to decline. Despite this performance dip at longer distances, the utilization of a greater number of RIS elements leads to improved data rate performance. These findings underscore the significance of an optimal number of RIS elements to attain enhanced data rate performance in long-distance communication scenarios. Careful consideration of the number of RIS elements in specific communication contexts can thus lead to more efficient and reliable data transmission.
\begin{figure}[t]
\centering
\begin{minipage}{0.5\textwidth}
                 \centering
                 \captionsetup{justification=centering,singlelinecheck=false}
                \includegraphics[width=\linewidth, height=2.2in]{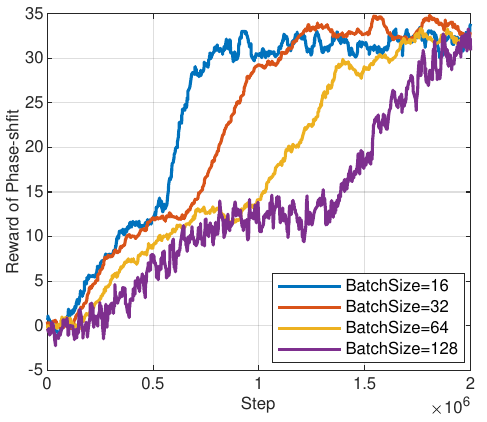}
                \caption{Phase-shift agent's convergence vs batch sizes.}
                \label{RIS_reward_vs_batchsize}
\end{minipage}%
\begin{minipage}{0.5\textwidth}
                \centering
                \captionsetup{justification=centering,singlelinecheck=false}
                \includegraphics[width=\linewidth, height=2.2in]{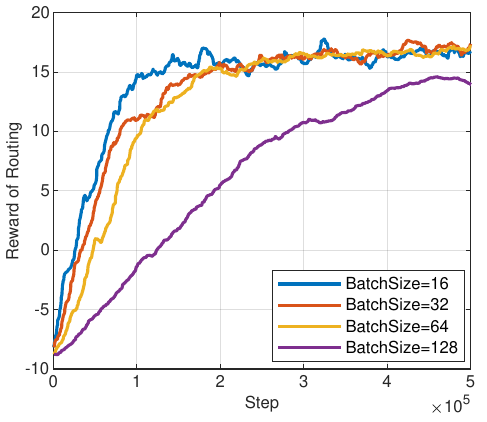}
                \caption{Routing agent's convergence vs batch sizes.}
                \label{Route_reward_vs_batchsize}
    \end{minipage}
      \vspace{-0.4in}
\end{figure}

Fig. \ref{RIS_reward_vs_batchsize} and Fig. \ref{Route_reward_vs_batchsize} illustrate the ablation study conducted on the RIS phase shift agent and the routing agent, respectively. The study investigates the impact of varying batch sizes on the rewards obtained by these agents, serving as an indicator of their learning performance. The findings indicate that as the batch size increases, the convergence of rewards becomes slower. This phenomenon is attributed to the efficiency of the PPO learning algorithm, which can adapt smoothly to lower batch sizes, as evident from the presented figures across different MDPs. The ablation study provides valuable insights into the interplay between batch sizes and learning performance for both the RIS phase shift and routing agents. Understanding these dynamics contributes to the optimization of learning processes in the context of the specific agents under consideration, thereby enhancing their performance in practical applications.
\vspace{-0.2in}
\section{Conclusion}
\label{conc}
\vspace{-0.1in}
This paper presents a novel approach to maximize LEO satellite coverage in 6G sub-THz networks by integrating RISs. The objective is to optimize network performance, including satellite-RUE association, data packet routing in satellite constellations, RIS phase shift, and GBS transmit power, to ultimately maximize end-to-end data rate. However, this optimization problem faces challenges due to the time-varying environment, non-convexity, and NP-hardness. To overcome these challenges, a BCD algorithm is proposed, combining BKMC, MAPPO-DRL, and the WOA. Simulation results show the proposed approach outperforms the baseline in network performance. This work advances satellite communication technologies for 6G by addressing challenges related to maximizing satellite coverage in sub-THz networks through RIS integration. Future work can explore semantic techniques \cite{chaccour2022less} to further enhance performance.
\ifCLASSOPTIONcaptionsoff
  \newpage
\fi
\vspace{-0.2in}
\bibliographystyle{IEEEtran}
\bibliography{TWC}
\end{document}